\documentstyle[aas2pp4,psfig]{article}

\received{soon} \accepted{soon after that} \journalid{}{}
\articleid{}{}

\slugcomment{to appear in A.J.}

\lefthead{Puzia et al.}  
\righthead{Age Differences between Globular
  Clusters in NGC~4472}

\begin{document}
\input{psfig.sty}

\title{The Age Difference between the Globular Cluster
  Sub--populations in NGC~4472}

\author{Thomas H. Puzia} \affil{Sternwarte der Universit\"at Bonn, Auf
  dem H\"ugel 71, D-53121 Bonn, Germany} \affil{UCO/Lick Observatory,
  University of California, Santa Cruz, CA 95064, USA}
\affil{Electronic mail: tpuzia@astro.uni-bonn.de}

\author{Markus Kissler-Patig \altaffilmark{1}} \affil{European
  Southern Observatory, Karl-Schwarzschild-Str.~2, D-85748 Garching,
  Germany} \affil{UCO/Lick Observatory, University of California,
  Santa Cruz, CA 95064, USA} \affil{Electronic mail:
  mkissler@eso.org}

\author{Jean P. Brodie \altaffilmark{2}} \affil{UCO/Lick Observatory, University of
  California, Santa Cruz, CA 95064, USA} \affil{Electronic mail:
  brodie@ucolick.org}

\author{John P. Huchra \altaffilmark{2}}  \affil{Harvard-Smithsonian Center for Astrophysics,
  60 Garden Street MS20, Cambridge, MA 02138-1516 USA} \affil{Electronic mail:
  huchra@cfa.harvard.edu}

\altaffiltext{1}{Feodor Lynen Fellow of the Alexander
  von Humboldt Foundation}

\altaffiltext{2}{Guest Observer, Hubble Space Telescope, operated by the
Association of Universities for Research in Astronomy for NASA}

\begin{abstract}
  
  The age difference between the two main globular cluster
  sub--populations in the Virgo giant elliptical galaxy, NGC~4472
  (M~49), has been determined using HST WFPC2 images in the F555W and
  F814W filters. Accurate photometry has been obtained for several
  hundred globular clusters in each of the two main sub--populations,
  down to more than one magnitude below the turn--over of their
  luminosity functions. This allows precise determinations of both the
  mean colors and the turn--over magnitudes of the two main
  sub--populations. By comparing the data with various population
  synthesis models, the age--metallicity pairs that fit both the
  observed colors and magnitudes have been identified. The metal--poor
  and the metal--rich globular clusters are found to be coeval
  within the errors ($\sim 3$ Gyr). If one accepts the validity of our
  assumptions, these errors are dominated by model
  uncertainties. A systematic error of up to 4 Gyr could affect this
  result if the blue and the red clusters have significantly different
  mass distributions. However, that one sub--population is half as old
  as the other is excluded at the 99\% confidence level. The different
  globular cluster populations are assumed to trace the galaxy's major
  star--formation episodes. Consequently, the vast majority of
  globular clusters, and by implication the majority of stars, in
  NGC~4472 formed at high redshifts but by two distinct mechanisms or
  in two episodes.
  
  The distance to NGC~4472 is determined to be $15.8\pm 0.8$ Mpc,
  which is in excellent agreement with six of the seven Cepheid
  distances to Virgo Cluster spiral galaxies. This implies that the
  spiral and elliptical galaxies in the main body of Virgo are at the
  same distance.

\end{abstract}



\section{Introduction}
\label{intro}
\subsection{Star formation in early--type galaxies traced 
  by globular cluster formation}

Arguably the two most important open questions about the formation of
early--type galaxies are when they formed their first stars and when
they assembled dynamically. While the latter question is probably best
answered by observations at intermediate and high redshift, the answer
to the former is likely to be found from observations in the more
local universe.

A number of studies have estimated the epoch of star formation by
comparison of observations of the diffuse stellar component in nearby
and low-redshift galaxies to stellar population models (e.g.~Renzini
1999 and Bender 1997 for reviews). However, the use of diffuse star
light has two main disadvantages. First, it is very difficult to
disentangle different stellar populations and, second, a relatively
recent but (in mass) unimportant star formation episode can dominate
the luminosity--weighted quantities used to derive the star formation
history.

Both these problems are by--passed when globular clusters are used to
study the main epochs of star formation. In nearby galaxies,
individual globular clusters appear as point source, and can be
characterized by a single age and a single metallicity. Distinct
populations can easily be identified (e.g.~in a color distribution of
the globular cluster system) and the mean properties of the
sub--populations can be determined. The relative numbers in each
sub--population will also indicate the relative importance of each star
formation episode.

There is strong support for the assumption that globular cluster
formation traces star formation. First, major star formation observed
today in interacting galaxies is accompanied by the formation of
massive young star clusters (e.g.~Schweizer 1997, although their exact
nature is still under debate, Brodie et al.~1998). Second, the
specific frequency (i.e.~number of globular clusters normalized to the
galaxy star light, see Harris \& van den Bergh 1981) is constant to
within a factor of two in almost all galaxies (see compilation in
Ashman \& Zepf 1998), which points to a link between star and cluster
formation. Moreover, although this relation seems to fail in giant
ellipticals, McLaughlin (1999) has shown that the number of globular
clusters is in fact proportional to the total mass, and the number per
unit mass is extremely constant from galaxy to galaxy. Finally, in
less violent star formation episodes, Larsen \& Richtler (1999) have
shown that the number of young star clusters in spirals is directly
proportional to the current star formation rate.

To first order, globular clusters are likely to be excellent tracers
for the star formation episodes in early--type galaxies.

\subsection{Several epochs of star formation}

There is now evidence for multiple globular cluster sub--populations
in a number of luminous, early--type galaxies (Zepf \& Ashman 1993,
Ashman \& Zepf 1998 for a recent compilation). The origin of these
sub--populations is still under debate (e.g.~Gebhardt \& Kissler-Patig
1999) but it seems clear that the main sub--populations formed in
different star--formation episodes, perhaps at different epochs. In
the best studied galaxies with multiple populations, the number of
clusters in each sub--population is roughly equal, suggesting that
these galaxies have had more than one {\it major} star--formation
episode. The overall evidence from photometry and more recently from
spectroscopy (Kissler-Patig et al.~1998a, Cohen et al.~1998) suggests
that in the cluster giant ellipticals the main sub--populations are
old. In other non--interacting early--type galaxies, the non--detection
of sub--populations, either due to their absence or to a conspiracy of
age and metallicity, leads to the same
conclusion: no major globular cluster (i.e.~star) formation has
occurred since $z\simeq1$ (Kissler-Patig et al.~1998b).

There have been a few previous attempts to determine the relative age
difference between globular cluster sub--populations. In the S0 galaxy
NGC~1380, Kissler-Patig et al.~(1997) derived an age difference of
around 3 to 4 Gyr between the halo and bulge globular clusters. In
M~87, Kundu et al.~(1999) estimated the metal--rich clusters to be 3
to 6 Gyr younger than the metal--poor ones. As a comparison, in the
Milky Way, the difference between halo and bulge/disk globular
clusters seems to be small (less than 1 Gyr: Ortolani et al.~1995,
around 17\%: Rosenberg et al.~1999).

Deriving relative ages from photometry alone is complicated by the
fact that broad band colors suffer from an age--metallicity degeneracy
(e.g.~Worthey 1994). One possible solution to this problem is to
measure several quantities that are affected differently by age and
metallicity and to combine the results. For example, most broad band
colors are more affected by metallicity than by age. On the other
hand, optical magnitudes for a given mass are more affected by age
than they are by metallicity. A combination of both can break the
age--metallicity degeneracy, when the mass distributions are known (see
Sect.~\ref{models}).

Both a mean color as well as a ``mean magnitude'' can be measured for
a globular cluster sub--population. The mean color is simply the mean
color of the globular cluster sub--sample, and can be determined from
a color distribution of the globular clusters. The turn--over (TO)
point of the globular cluster luminosity function (GCLF) of a given
sub--population can be used as its ``mean magnitude''. The absolute TO
magnitude of the GCLF has been found to be fairly constant from one
globular cluster system to the next in a wide variety of galaxy types
and environments with a peak at $M_V^{TO}\simeq -7.5$ (see Whitmore
1997 and Ferrarese et al.~2000 for summaries). Three quantities affect
the TO value: age, metallicity and mass. The effects of age and
metallicity on the magnitude of a single stellar population were
quantified by Ashman, Conti \& Zepf (1995) using population synthesis
models from Bruzual \& Charlot (1995). The mean mass influences the TO
magnitude because the TO of the GCLF corresponds to a break in the
globular cluster mass distribution (e.g.~McLaughlin \& Pudritz 1996).
That is, if the characteristic mass (at which the slope of the power
law mass function changes) varies, the TO magnitude will be affected.
However, the mass distributions seem to be extremely similar from
galaxy to galaxy and are expected to be very similar for different
sub--populations within the same galaxy if destruction processes are
the dominant mechanism for shaping the mass distribution, and had time
enough to act on both sub--populations. We will come back to this
point in Section \ref{discussion}.

Neither the mean color nor the TO magnitude alone can be used to
derive the age and metallicity of a globular cluster population
because of the above--mentioned degeneracy. This paper marks the first
time that both the mean color {\it and} the TO magnitude of each
globular cluster sub--population have been simultaneously compared to
single stellar population (SSP) models (see Sect.~\ref{models}) in
order to distinguish age and metallicity. Deep high--quality
photometry is required to reliably determine the mean color and the TO
magnitude (implying good photometry below this TO point) and clean
globular cluster samples are needed, i.e.~good discrimination between
globular clusters and fore--/background contamination is essential.
The Wide Field and Planetary Camera 2 (WFPC2) on--board the Hubble
Space Telescope (HST) provides the most accurate photometric data
available and makes such an analysis feasible.

\subsection{Globular clusters in NGC~4472}

For this first study, we selected a galaxy that hosts a populous
globular cluster system which provides good number statistics even
after the sample is split into the different sub--populations.

NGC~4472 (M~49), the brightest giant elliptical galaxy in the Virgo
cluster, is located in the center of a sub--group within the cluster
and is of Hubble type E2. It is one of the brightest early--type
galaxy within 20 Mpc with an absolute magnitude of $M_V\simeq-22.6$
(See Table 1 for other characteristics). It hosts $6300\pm 1900$
globular clusters (Geisler et al.~1996). The specific frequency of
globular clusters is $S_N=5.6\pm 1.7$, close to the average for bright
cluster ellipticals.

Previous photometric studies of the globular cluster system of this
galaxy include those of Cohen (Gunn--Thuan photometry, 1988), Couture
et al.~(Johnson photometry, 1991), Ajhar et al.~(Johnson photometry,
1994), and Geisler et al.~(1996) and Lee et al.~(1998, both Washington
photometry).

The two most recent studies were the first to obtain a comprehensive
enough data set to allow the identification and characterization of
the two main globular cluster sub--populations. The mean metallicities
of the two sub--populations were found to be [Fe/H]$\simeq-1.3$ and
$-0.1$ dex for the blue and red peak, respectively. The observed
metallicity gradient ($\Delta$[Fe/H]/$\Delta$ log$r \simeq -0.4$
dex/log(arcsec)) in the system is mostly due to the radially varying
ratio of these two populations. The metal--rich component is spatially
more concentrated and has the same ellipticity as the galaxy star
light while the metal--poor component is more extended and spherical.
These two components can be loosely associated with the ``halo'' and
``bulge'' of NGC~4472 suggesting different formation mechanisms for
these galaxy structures and the associated clusters.  The relative
ages of these two components is the subject of this paper.

The paper is structured as follows: In Section \ref{observation} we
present the observation and data reduction. Section \ref{analysis}
contains details of the data analysis; color and turn--over magnitude
determinations, etc. Section \ref{agediffs} compares the results to
stellar evolutionary models to determine the age difference between
the sub-populations. In section \ref{distance} we derive a new
distance to NGC~4472 using the TO magnitude method. Section
\ref{summary} summarizes the results.


\section{Observations and Data Reduction}
\label{observation}

\subsection{Observations}

The WFPC2 images of NGC~4472 were obtained from two data sets. The
nuclear pointing was taken from Westphal et al.~(from HST program
GO.5236) and was complemented by our own pointings to the north and
south of the galaxy (HST program GO.5920, PI: Brodie) which were
offset in DEC by 2\arcmin\ 16\arcsec\ and 2\arcmin\ 49\arcsec\ from
the center of the galaxy, respectively. The data are summarized in
Tab.~2.

\subsection{Data Reduction}

\subsubsection{Image processing}

We used calibrated science images returned by the STScI\footnote{Based
  on observations with the NASA/ESA Hubble Space Telescope, obtained
  from the data Archive at the Space Telescope Science Institute,
  which is operated by the Association of Universities for Research in
  Astronomy, Inc.~under NASA contract No.~NAS5-26555.}. The basic
reduction was carried out in {\tt IRAF}\footnote{{\tt IRAF} is
  distributed by the National Optical Astronomy Observatories, which
  is operated by the Association of Universities for Research in
  Astronomy, Inc., under cooperative agreement with the National
  Science Foundation.}. The images were combined with the task {\tt
  crrej} of the {\tt STSDAS} package. The Brodie data set consists of
paired, dithered exposures, and was thus registered by shifting the
second of each image pair by 0.506\arcsec\ (i.e.~5 pixels for the WF
chips, 12 pixels for the PC chips) with the task {\tt imshift}.

Using the standard aperture radius of Holtzman et al.~(1995) of
0.5\arcsec , we tested for any magnitude error that might be
introduced by incomplete shifting of frames. This error was found to
be less than 0.0003 mag for both filters and is therefore negligible.

\subsubsection{Photometry}

The photometry was carried out using the source extraction software
{\tt SExtractor v2.0.19} by Bertin \& Arnouts (1996). For the purpose
of object selection only, the images were convolved with the
instrumental PSF to improve the detection efficiency. The PSF for each
filter was taken from the HST Handbook version 4.0, using the PSFs
(pixel centered) at 600 nm for F555W and at 800 nm for F814W. The
selection criteria were set to 2 connected pixels, 2$\sigma$ above the
background computed in a $32\times32$ pix$^2$ grid. Magnitudes were
measured in 2 pixels and 0.5\arcsec\ radius apertures (see also next
Section). Objects in common on the $V$ and $I$ frames were identified
and matched. The positions on the sky of all objects were computed
using the task {\tt metric} in the {\tt STSDAS} package.

The calibration and transformation to the Johnson--Cousins $V$ and $I$
filters followed Holtzman et al.~(1995) and corrections for charge
transfer efficiency and gain ($7e^-/ADU$ for all our data) were
applied. In what follows, the magnitude errors will {\it not} include
the $<4\%$ systematic error potentially introduced by the calibration.
The values measured in the 0.5\arcsec\ radius apertures were used to
compute the final magnitudes in order to avoid having to apply an
additional correction to the data.

\subsubsection{Extended objects calibration}
\label{extended}

Globular clusters at the distance of the Virgo galaxy cluster are
expected to be resolved. Milky Way globular clusters have typical
half--light radii between 3 and 10 pc (e.g.~Harris 1996),
corresponding to 0.04\arcsec\ to 0.13\arcsec\ at 16 Mpc distance.
Indeed, most of our globular cluster candidates have FWHM measurements
(as returned by {\tt SExtractor}) slightly larger than measured for
unresolved objects. This, in turn, implies that the photometric
corrections proposed by Holtzman et al.~(1995) for unresolved objects
will not be valid for converting from the 0.5\arcsec\ radius aperture
to the total globular cluster magnitude.

New rough corrections were derived as follows. We computed
$30\times30$ pix$^2$ normally sampled PSFs in $V$ and $I$ for a WF
chip using the {\tt Tiny Tim v4.4} program by Krist \& Hook (1997),
and convolved these PSFs with Hubble laws of different core radii
(0.05\arcsec, 0.1\arcsec, 0.2\arcsec, 0.6\arcsec). Note that these
PSFs include the convolution with the diffusion kernel (cf.~Krist \&
Hook 1997) which is not well defined. This kernel introduces an
intrinsic smoothing to the optical PSF (as seen at the top of the CCD
chip) due to charge diffusion.

The cumulative light distributions of these modeled objects were
compared with those of a set of isolated globular cluster candidates
of various luminosities and chip positions taken from our data. The
data in $I$ match almost perfectly the PSF convolved with a Hubble law
of 0.1\arcsec\ core (corresponding to $\sim 8$ pc at a distance of 16
Mpc), while the $V$ data lie in between the PSF models convolved with
0.05\arcsec\ and 0.1\arcsec\ Hubble laws.

The same procedure, using 10 times over--sampled PSFs without applying
any diffusion kernel, yielded best fitting Hubble--core radii of
0.3\arcsec\ for $I$ data and 0.1\arcsec\ for the $V$ data. Until the
behavior of the diffusion kernel is better understood (wavelength
dependencies, etc.), it is possible that a systematic core--radius
error of about 0.1\arcsec\ ($\sim 8$ pc) may be present in all radius
estimation studies of Virgo globular clusters using WFPC2 data.
Nonetheless, we used the normally--sampled and diffusion--kernel
convolved PSFs for our analysis.

\begin{figure}
  \psfig{figure=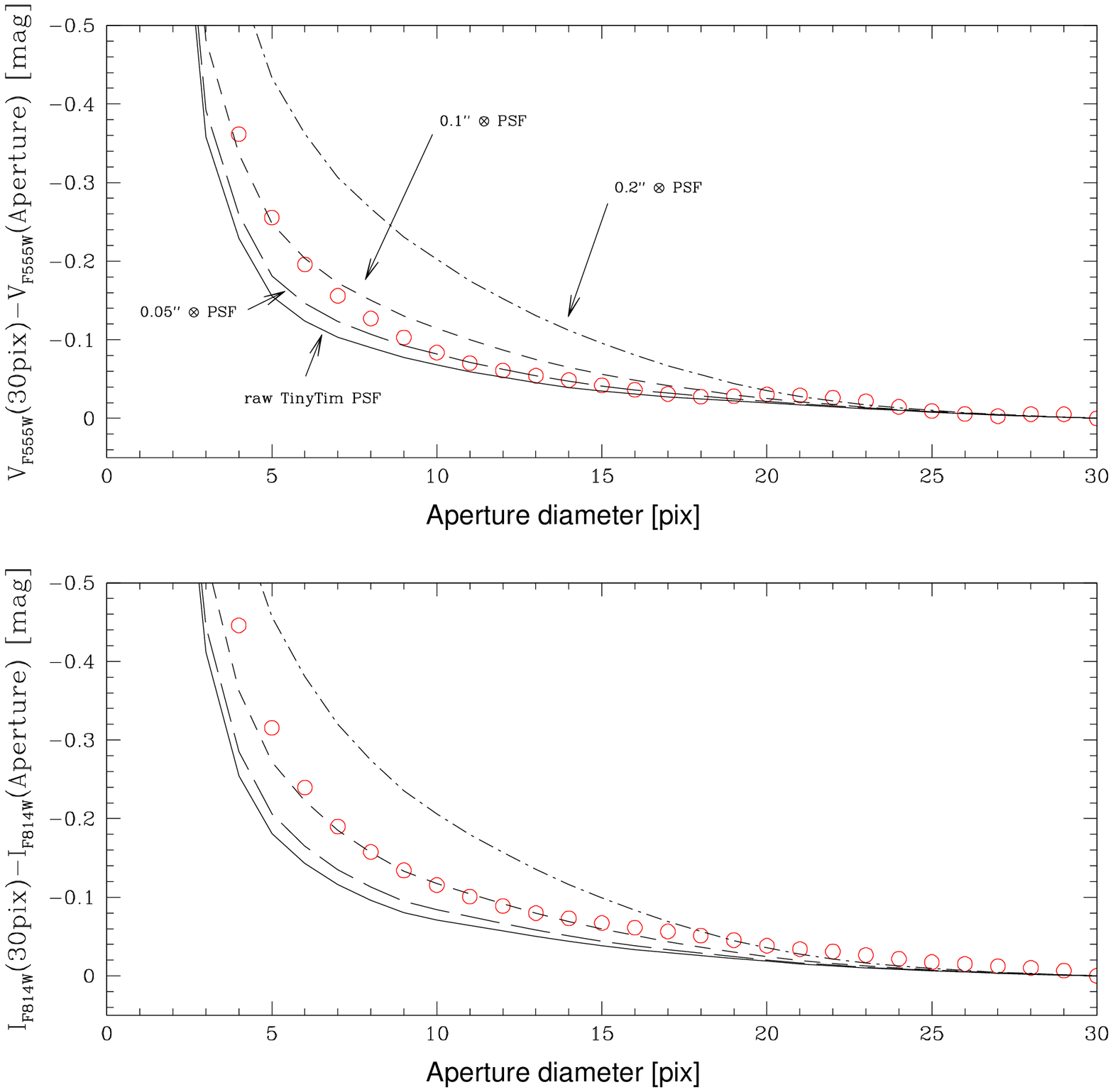,height=16cm,width=16cm
    ,bbllx=8mm,bblly=57mm,bburx=205mm,bbury=245mm}
\caption{Curves of growth for PSF models and selected data. The curves
  represent different models (solid line = raw {\tt TinyTim} PSF;
  long--dashed line = Hubble law (0.05\arcsec\ core radius) $\otimes$
  PSF; short--dashed line = Hubble law (0.1\arcsec\ core radius)
  $\otimes$ PSF; dot--dashed line = Hubble law (0.2\arcsec\ core
  radius) $\otimes $ PSF while the open circles show the selected
  globular cluster data: 13 candidates in $V$ and 15 in $I$.}
\label{ps:wachstum}
\end{figure}

The modeled light profiles (see Fig.~\ref{ps:wachstum}), unlike the
real data, provide high enough signal--to--noise to determine the
extrapolation from an 0.5\arcsec\ radius to the total $V$ and $I$
magnitudes. The derived corrections are applied to the 0.5\arcsec\ 
radius aperture measurements in addition to the Holtzman point source
correction of $C=0.1$. These additional corrections are $C_V =
0.050\pm0.015$ and $C_I = 0.080\pm0.010$ (which results in a color
correction of $C_{(V-I)} = 0.030\pm0.018$). The errors reflect the
uncertainty in matching the data with the models and the accuracy of
the extrapolation to infinity of our fit to the light profile. The
errors do not include, for example, the fact that the core and the
half--light radii of different globular clusters may vary
significantly. In general, our data were calibrated applying the
correction for point sources of $C = 0.10$. When additional
corrections were included, they are specifically mentioned in the
text. Note that the corrections are systematic, i.e.~they do not
affect {\it relative} magnitudes. Therefore, the accuracy with which
we are able to determine age--metallicity {\it differences} should not
be seriously affected.

\subsubsection{Artificial star/cluster experiments}
\label{artstar}

Artificial star/cluster experiments were carried out using the tasks
{\tt starlist} and {\tt mkobjects} from the {\tt IRAF} package, {\tt
  artdata}. 1000 objects were added per run in both the $V$ and $I$
images. 400 runs were computed resulting in 400,000 artificial objects
per filter and covering the whole parameter space of background
values. Both objects modeled by a WF chip PSF and objects modeled by
the same PSF but convolved with Hubble laws of core radius
0.05\arcsec\ ($V$) and 0.1\arcsec\ ($I$) were added. All objects had a
color of $V-I=1.1$, previous tests with 5000 objects having shown no
significant variations among the completeness curves for $V-I=$~0.9,
1.1 and 1.3.

\begin{figure}
  \psfig{figure=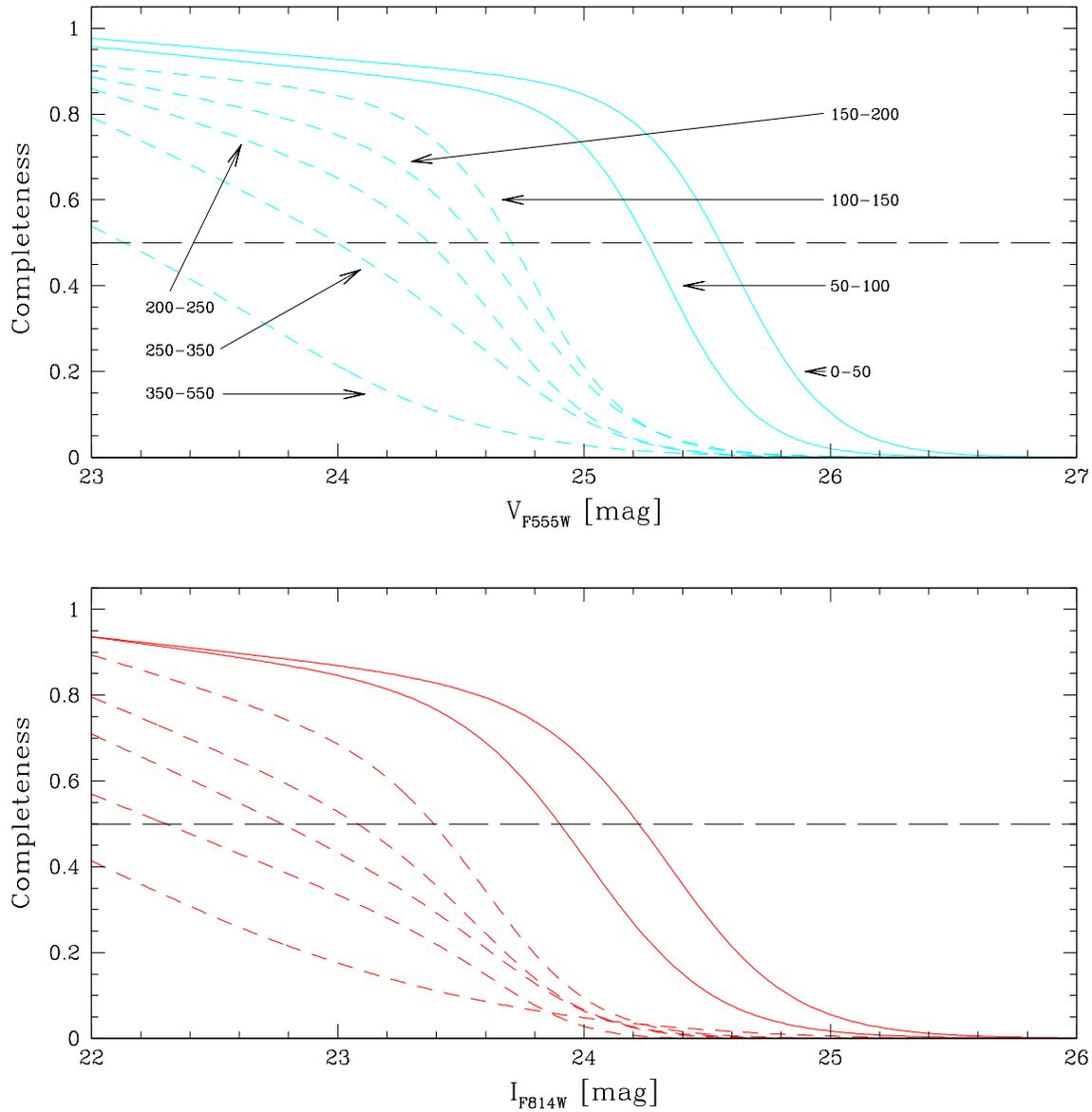,height=16cm,width=16cm
    ,bbllx=8mm,bblly=57mm,bburx=205mm,bbury=245mm}
\caption{Completeness curves for the central pointing (dashed
  lines) and for the north-- and south--pointings (solid lines). The
  central images have much higher background values than the
  northern and southern pointings. Each curve refers to a particular
  background value bin, i.e.~from faint to bright magnitudes in
  counts: 0-50, 50-100, 150-200, 200-250, 250-350, 350-550.}
\label{ps:inc}
\end{figure}

The selection completeness was computed for objects found on both the
$V$ and $I$ images. All detected objects which passed our selection
criteria (see Sect.~\ref{selection}) and all input objects were sorted
by their background value and subdivided in 10 background--value bins
for $V$ and $I$, respectively. A {\it skewed Fermi function}
\begin{equation}
F(x)=\frac{ax+b}{1+\exp(x-c)^d}
\end{equation}
was fit to all completeness histograms for each background bin. A
smoothing procedure using a low--pass filter gives nearly the same
results. The 50\% completeness limits lie around $V\simeq 25.5$ and
$I\simeq 24.5$ for the lowest background value. The 50\% completeness
magnitude falls off continuously with a growing background (see
Fig.~\ref{ps:inc}).

The artificial--object experiments also allowed us to determine the
error in the photometry and to test the accuracy with which the FWHM
of objects can be recovered as a function of magnitude. Typical FWHM
values were deduced for extended objects which are a good match to the
FWHM values of our globular cluster candidates.

\subsubsection{Globular cluster selection criteria}
\label{selection}

\begin{figure}
  \psfig{figure=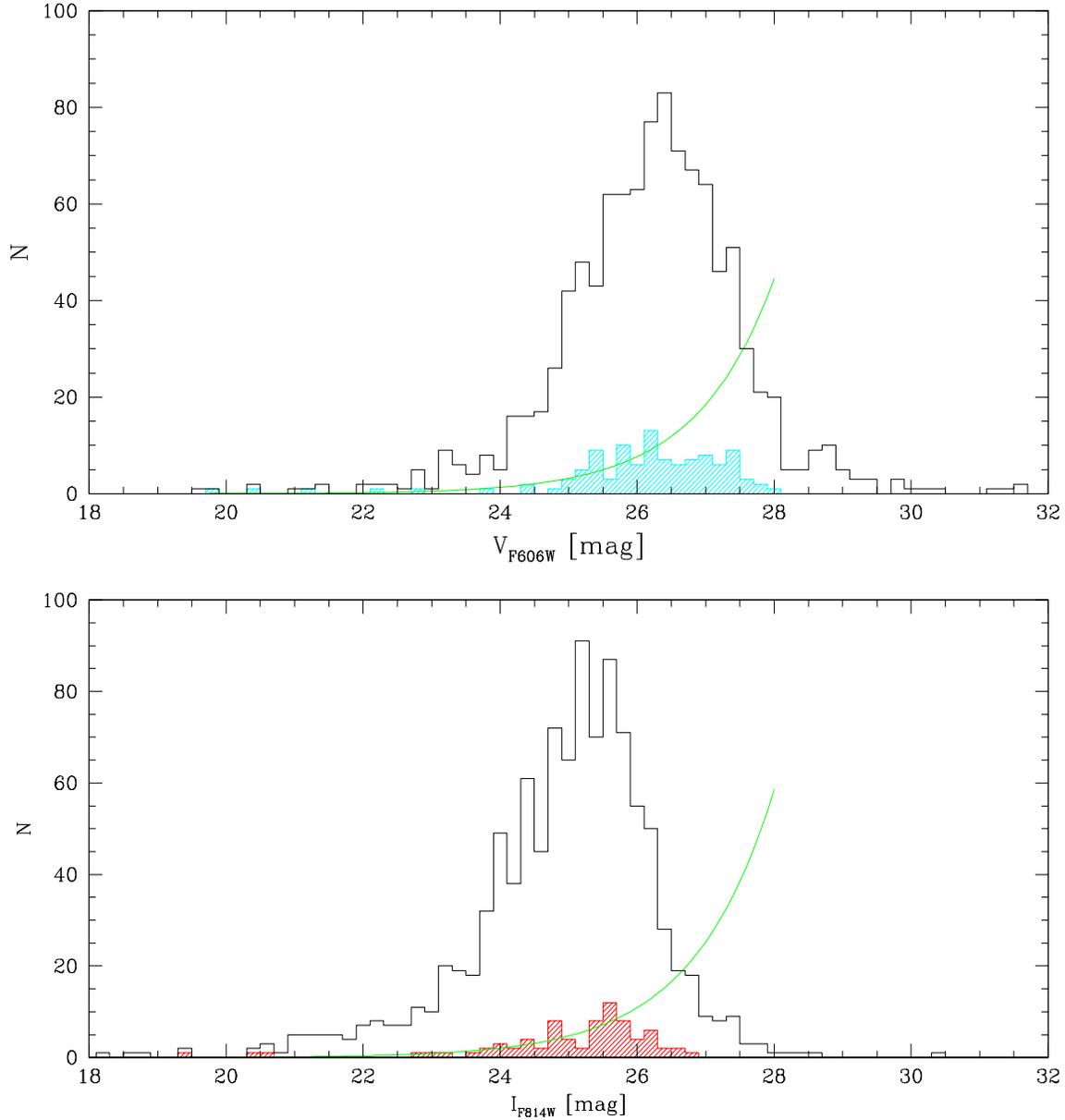,height=16cm,width=16cm
    ,bbllx=8mm,bblly=57mm,bburx=205mm,bbury=245mm}
\caption{Background contamination histograms for the $V_{F606W}$ 
  and $I_{F814W}$ band resulting from the HDF--N. The unshaded
  histograms show the background object luminosity distribution as
  obtained from HDF--N directly, while the shaded ones show the same
  data after applying our criteria. The lines are exponential least
  square fits to the small histograms. In order to ensure a reasonable
  fit we set the upper limit for the magnitude range at the 50\%
  completeness level quantified during the completeness tests. Using
  the exponential law, the integrated number of background objects is
  $\sim 30$ up to the 50\% completeness limit in both filters.}
\label{ps:back}
\end{figure}

Object selection criteria were applied to our extracted data to
provide a clean sample of globular cluster candidates for further
analysis. Only objects which met the color cut of $0.5 < V-I < 1.8$
and a magnitude cut of $V\leq$ 25.5 and $I\leq$ 25.2 were included so
that some of the red unresolved galaxy nuclei were rejected
(cf.~Fig.~\ref{ps:back}). In addition, a FWHM cut in the range 0.75
pix $<$ FWHM $<$ 2.5 pix was applied. We also calculated the magnitude
difference obtained through apertures of 0.5 pix and 3 pix radii which
allowed us to distinguish between point--source objects and extended
objects like globular clusters (e.g.~Schweizer et al.~1996). A very
rough upper cut of $\Delta(0.5 - 3)>1.0$ was used.

The neural network of {\tt SExtractor v2.0.19}, i.e.~the $CLASS$
parameter was used to supplement the above selection criteria. This
parameter was defined for well--sampled images (Bertin \& Arnouts
1996). Since the WFPC2 provides under--sampled images the reliability
of the $CLASS$ parameter is reduced. We applied only the selection
criterion, $CLASS < 1.0$, to our data. A test showed that for objects
brighter than the 50\% completeness level the $CLASS$ parameter cut
reliably eliminates point sources and cosmic rays.

Our final list of globular cluster candidates includes 705 objects in
$V$ and $I$. This is 38\% of the initially extracted sample. An
electronic list of the globular--cluster candidates is available from
the authors.

\subsubsection{Background contamination}

One significant advantage of HST data over ground--based observations
is high spatial resolution which allows high--confidence rejection of
extended objects. However, background contamination is still expected.
In order to determine the background fraction in our data we ran the
same extraction and selection criteria for the Hubble Deep Field North
(HDF--N, Williams et al.~1996) images. Since the HDF--N has galactic
coordinates of $l=$ 125.89 and $b=$ +54.83 while our exposures have
$l=$ 286.92 and $b=$ +70.20, we might overestimate the contamination
due to foreground stars using this approach, although foreground star
contamination is, in any case, expected to be negligible. We obtained
from the HST Archive images in F606W and in the F814W filter with
exposure times of 20200 sec for F814W and 15200 sec for F606W,
respectively. These images are far deeper than our own. Both
magnitudes for F606W and F814W were transformed to Johnson $V$ and
Cousins $I$ magnitudes. We determined contamination histograms with
and without the application of our selection criteria in order to
establish their reliability.  Exponential laws have been fit to the
histograms, after applying the rejection criteria, up to the 50\%
completeness limits. The exponential laws act as input for our
Improved Maximum--Likelihood Code (see Sect.~\ref{TOmags}).

Note the efficiency of our selection criteria. Nearly all background
objects are removed from the data up to the 50\% completeness
magnitude. The contamination is less than 5\% for both the $V_{F606W}$
and $I_{F814W}$ filters (see Fig.~\ref{ps:back}).

We stress again that all the above corrections were applied to the
full sample, i.e.~differential values between sub--samples remain
unaffected even if small systematic effects have been introduced.


\section{Analysis}
\label{analysis}

Our goal is to determine the mean colors and turn--over magnitudes of
the luminosity functions of the two main globular cluster
sub--populations in both the $V$ and the $I$ filters. These colors and
magnitudes can then be compared to the predictions of population
synthesis models for given ages and metallicities.

\subsection{Mean colors}
\label{meancol}
\begin{figure}
  \psfig{figure=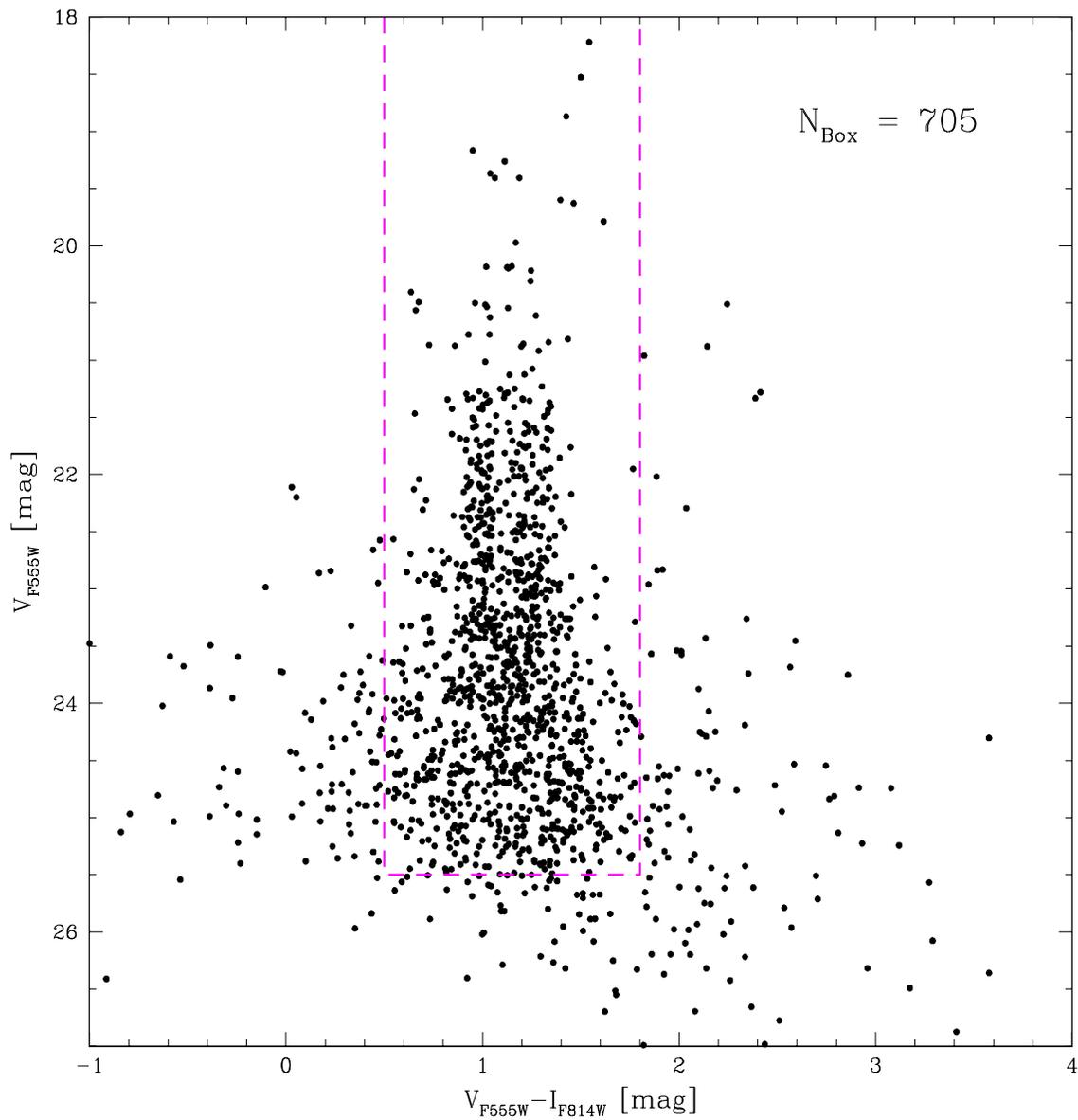,height=16cm,width=16cm
    ,bbllx=8mm,bblly=57mm,bburx=205mm,bbury=245mm}
\caption{Color magnitude diagram of globular clusters in NGC~4472.
  The box delineates the initial cuts made to the raw data prior
  analysis. 705 globular cluster candidates are enclosed. }
\label{ps:fhd}
\end{figure}

\begin{figure}
  \psfig{figure=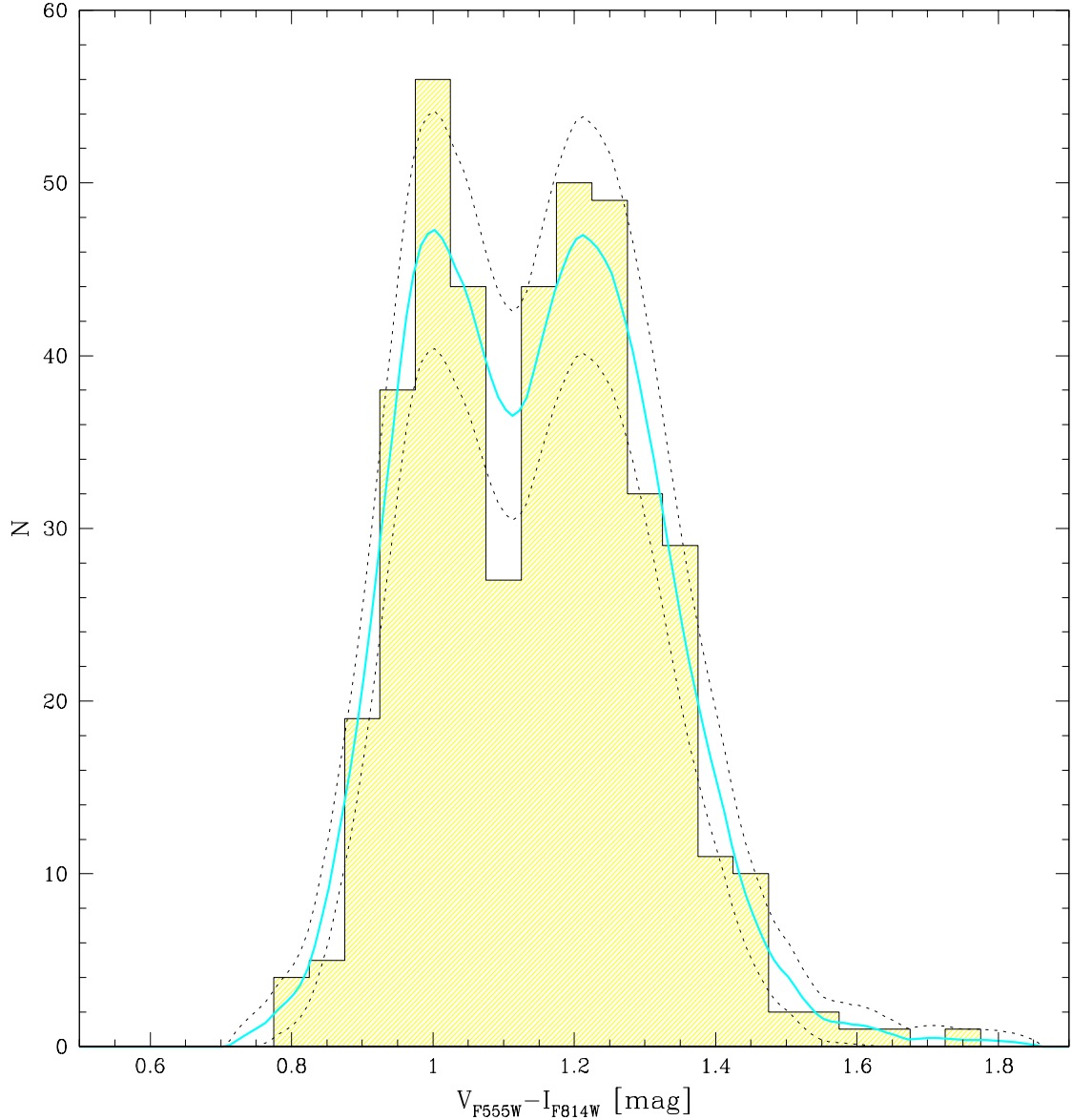,height=16cm,width=16cm
    ,bbllx=8mm,bblly=57mm,bburx=205mm,bbury=245mm}
\caption{Color distribution of globular clusters in NGC~4472. A bimodal
  distribution is found to be present at the $99.99$\% confidence
  level using the KMM mixture--modeling algorithm (Ashman et
  al.~1994). The two peaks lie at $0.99 \pm 0.01$ and $1.24 \pm 0.01$
  ($0.93 \pm 0.01$ and $1.18 \pm 0.01$ corrected for reddening and
  aperture). A further analysis using a density estimation technique
  (kernel estimator with an Epanechnikov kernel) yields color peaks at
  $V-I=1.00\pm 0.01$ and $1.22\pm 0.01$. The solid line depicts the
  derived distribution function with statistical upper and lower error
  limits (short--dashed lines).}
\label{ps:color}
\end{figure}

In order to determine the mean colors of the sub--populations we
selected from the total sample objects with small photometric errors.
The application of a magnitude cut at $V=23.75$ resulted in errors
$\Delta(V-I)<0.05$. We further applied an upper magnitude cut at
$V=20.0$ (corresponding to $M_V\simeq -11.2$ at the distance of Virgo)
in order to exclude any foreground star contamination, and a broad cut
in color ($0.8<(V-I)<1.8$) to isolate the region where globular
clusters are expected. The color--magnitude diagram of our sample is
shown in Fig.~\ref{ps:fhd}, together with the box (dotted lines) of
pre--selected objects (see Sect.~\ref{selection}). A $(V-I)$ histogram
of those selected objects is shown in Fig.~\ref{ps:color}.

Formally, a key underlying assumption is that the color (metallicity)
does not significantly vary with magnitude (mass) within a
sub--population. This appears to be the case as far as can be tested
with our data.

A KMM test (Ashman, Bird \& Zepf 1994) was run on our constrained
sub--sample. The test determines the confidence with which bimodality
can be said to be present, the most likely relative contribution of
each population to the sample, and the peaks (i.e.~mean colors) of
each sub--population. We find that the color distribution is bimodal
at the $>99.99$\% confidence level. The relative contributions of
objects to the selected sample (cf.~Sect.~\ref{selection}) are 282 and
423 in the blue and red populations respectively. The peaks of the
color distribution lie at $(V-I) = 0.99 \pm 0.01$ and $1.24 \pm 0.01$
for the blue and red sub--populations, respectively. Formally, the
peaks can be determined with an accuracy of 0.005 mag. The error
adopted here reflects the range of values derived by varying the
selection criteria (i.e.~magnitude and color cuts).

Two fitting modes were used: 1) forcing identical dispersions
(homoscedastic fit), 2) allowing independent dispersions
(heteroscedastic fit). We note that substantially better fits, in
terms of reduced $\chi^2$, are obtained if the dispersions of the blue
and red distributions are allowed to differ, and that the red
population seems to have a color distribution twice as broad as the
blue one ($\sigma_{\rm blue}\simeq 0.06$, $\sigma_{\rm red}\simeq
0.12$).

We estimated the peak colors with yet another method that avoids the
assumption of a Gaussian distribution for each color peak (intrinsic
to the KMM code). We computed the density distribution in color of the
objects with the help of a kernel estimator using the Epanechnikov
kernel (e.g.~Gebhardt \& Kissler-Patig 1999). We used $h_{\rm
  opt}=0.043$ for the smoothing parameter (see Silverman 1986 for
detailed discussion), an average of the optimal smoothing parameters
for the blue and red peak ($h_{\rm opt}(\mbox{blue})=0.023$ and
$h_{\rm opt}(\mbox{red})=0.063$) estimated from the $\sigma$--guess of
KMM for both modes. The density distribution is overplotted in
Fig.~\ref{ps:color}. A Biweight fit to each peak returns $(V-I)=
1.00\pm0.01$ and $1.22\pm0.01$, in perfect agreement with the results
from the KMM code.

To obtain the mean colors of the peaks, the above $(V-I)$ values need
to be corrected for the calibration offset described in
Sect.~\ref{extended} ($\Delta(V-I)=0.030\pm0.018$), and for a Galactic
reddening towards NGC~4472 of E$(B-V)=0.0224$ (Schlegel et al. 1998),
i.e.~E$(V-I)=1.3\:\cdot\:{\rm E}(B-V)=0.029$ (following Dean et
al.~1978). Note that the {\it relative} color difference, later used
to derive the age difference, is only affected by these corrections in
second order. The final, corrected mean colors of the two populations
are $(V-I)_0=0.93\pm0.02$ and $(V-I)_0=1.18\pm0.02$.

\subsection{Turn--over magnitudes}

The TO magnitudes of the luminosity functions for the blue and red
sub--population were individually determined in each filter using a
Maximum Likelihood approach (cf.~Secker 1992). The changing background
counts over our whole sample motivated us to develop a version of
Secker's (1992) Maximum Likelihood Estimator which includes background
variations. We calculated the completeness as a function of magnitude
and background noise as described in Sect.~\ref{artstar}. This
improved code, described below, was used to derive the peak of the
magnitude distribution for each filter and each sub--population.

Secker (1992) showed that the GCLF of the Milky Way and M31 was best
represented analytically with the Student's $t_5$-function
\begin{equation}
  \label{eq:t5}
  t_5(m|m_o,\sigma_t) = 
  \frac{8}{3\sqrt5\pi\sigma_t}
  \left(1+\frac{(m-m_o)^2}{5\sigma_t^2}\right)^{-3}.
\end{equation}
We adopted the Student's $t_5$--function as the most appropriate
distribution for the GCLF of NGC~4472 since it matches the tails of
the distribution better than a Gaussian function. The difference in
the result using the one or the other function is, however, very
small.

\subsubsection{Improved Maximum Likelihood}
\label{TOmags}
Usually a dataset of $n$ independent observations, like a sample of
magnitudes $M=(m_1,m_2,...,m_n)$, is distributed in an unknown way.
Assuming a distribution function $\phi(M|\Omega)$ which describes the
distribution of data properly and depends on $q$ parameters
$\Omega=(p_1,p_2,...,p_q)$ (particularly the turn--over magnitude
$m_o$ and the dispersion $\sigma_t$) the likelihood function $l(\Omega
|M) = \phi(M|\Omega)$ is considered as a function $l$ of $\Omega$ for
a given dataset $M$. The likelihood function is
\begin{equation}
 \label{eq:ml}
 l(\Omega |M) = \prod\limits_{i=1}^n \phi(m_i|\Omega).
\end{equation}
For convenience, during the calculation the logarithmic likelihood
function is used. During the evaluation the best set of parameters
$\bar{\Omega}$ is found if
\begin{equation}
 \label{eq:maxl}
 \frac{\partial l(\Omega)}{\partial\Omega}|_{\Omega = \bar{\Omega}}=0.
\end{equation}
For a detailed treatment of Maximum Likelihood theory see, for
instance, Bevington \& Robinson (1992).

Our improved Maximum Likelihood code, based on Secker's (1992) code,
takes into account the influence of the background--noise level $\rho$
on the completeness function $I(m,\rho)$, in addition to the
magnitude. Similarly, the photometric error function
$\varepsilon(m,\rho)$ now also depends on these two variables.  The
distribution function is
\begin{equation}
  \label{eq:distribution}
    \begin{array}{l}
    \phi(m|m_o,\sigma_t,\rho)=\\
    \quad K\cdot\Phi(m|m_o,\sigma_t,\rho)
    +(1-K)\cdot B(m|\rho)
  \end{array}
\end{equation}
\noindent 
where $m_o$ is the turn--over magnitude of the given distribution and
$\sigma_t$ is its dispersion (note: $\sigma_{\rm
  Gauss}=1.291\cdot\sigma_{t_5}$), while $\rho$ is the background
noise for the evaluation position on the chip. $K=N_{\rm GC}/(N_{\rm
  GC}+N_{\rm HDF})$ is the mixing parameter, i.e.~the fraction of
globular--cluster candidates in the whole data set including the
background objects. Note that, in our case, K is close to 1.

It is necessary to normalize all the functions used, which is
implicitly accomplished by
\begin{equation}
  \label{eq:Phi}
  \begin{array}{l}
    \Phi(m|m_o,\sigma_t,\rho)=\\
    \quad I(m|\rho)\cdot\left(\int
      t_5(m'|m_o,\sigma_t)\cdot
      \varepsilon(m|m',\sigma_t)\,dm' \right)
  \end{array}
\end{equation}
and
\begin{equation}
  \label{eq:beta}
  B(m|\rho)=I(m|\rho)\cdot b(m)
\end{equation}
\noindent 
where $b(m)$ is the background contamination function obtained from
the HDF data. $\varepsilon$ is the photometric error function
\begin{equation}
  \label{eq:photerr}
  \varepsilon(m|m',\rho)=\frac{1}{\sqrt{2\pi}\cdot\sigma(m',\rho)}
  \cdot\exp\left(\frac{(m-m')^2}{2\cdot\sigma(m',\rho)^2}\right)
\end{equation}
\noindent 
which is assumed to introduce Gaussian errors to the distribution
function $\phi$ with a magnitude and background noise dependent
dispersion
\begin{equation}
  \label{eq:error}
  \sigma(m,\rho)=\sqrt{10^{\:0.4\cdot(m-a)}
    +\rho^2\cdot 10^{\:0.8\cdot(m-b)}}.
\end{equation}
The parameters $a$ and $b$ can be estimated with sufficient accuracy
from a two--dimensional fit to the photometric errors obtained from
the artificial star experiments. They act as input parameters to our
code.

The need to account for the variable background, i.e.~the need for the
improved code, is best seen in Fig.~\ref{ps:inc}. The selection
completeness as a function of magnitude is plotted for a few typical
values of background counts found in our datasets. The 50\%
completeness limits shift to brighter magnitudes with increasing
background, i.e.~photon noise.

Alternatively one could approximate a ``mean'' completeness by
distributing the artificial stars with the same density profile as the
globular clusters, without including the background as a parameter in
the Maximum Likelihood analysis. The approach we have adopted is the
more accurate solution to the problem. The package described above is
available as an {\tt IRAF} task from the authors.

\subsubsection{Maximum Likelihood estimations}

\begin{figure}
  \psfig{figure=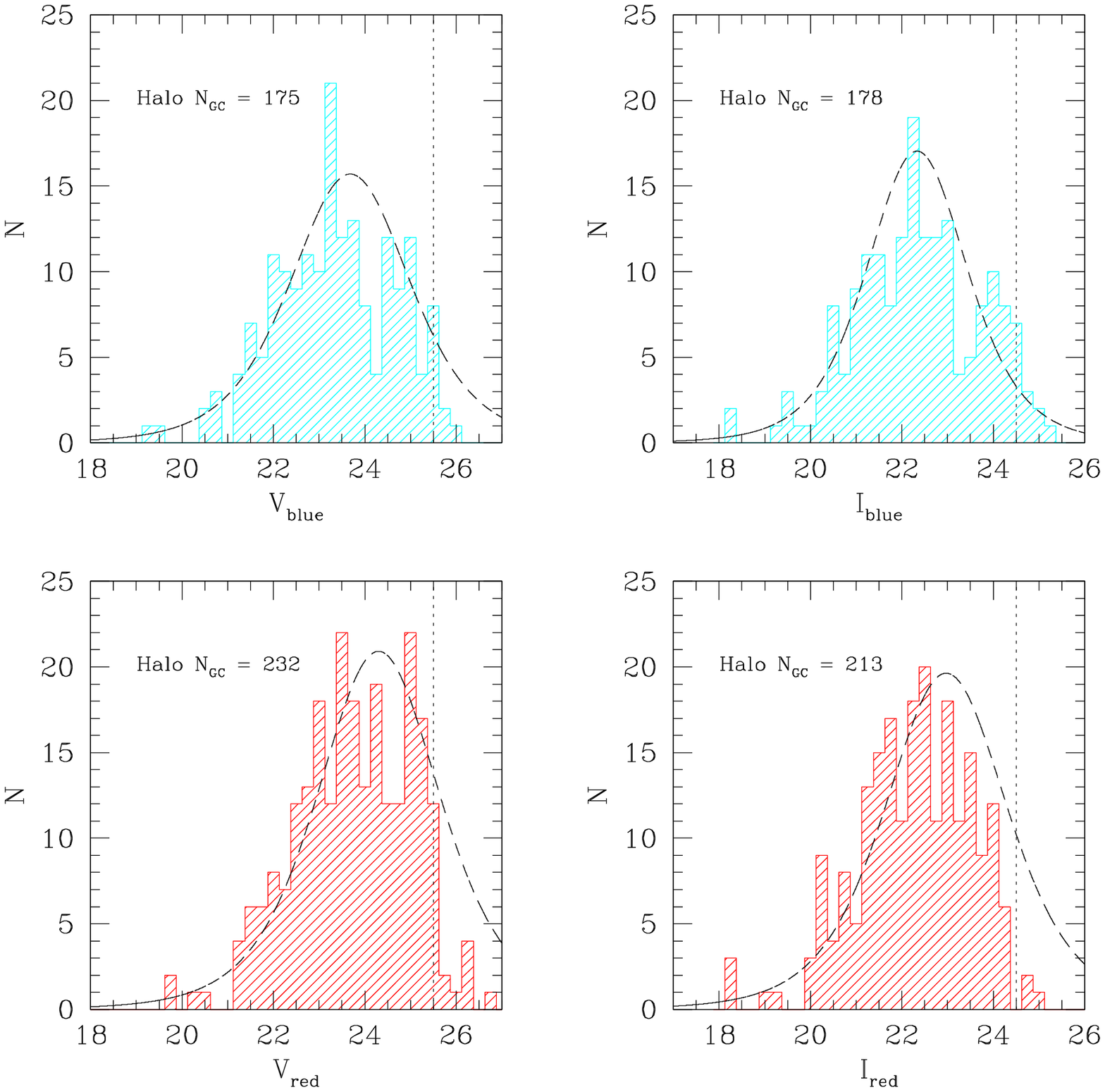,height=8cm,width=8cm
    ,bbllx=8mm,bblly=57mm,bburx=205mm,bbury=245mm}
  \psfig{figure=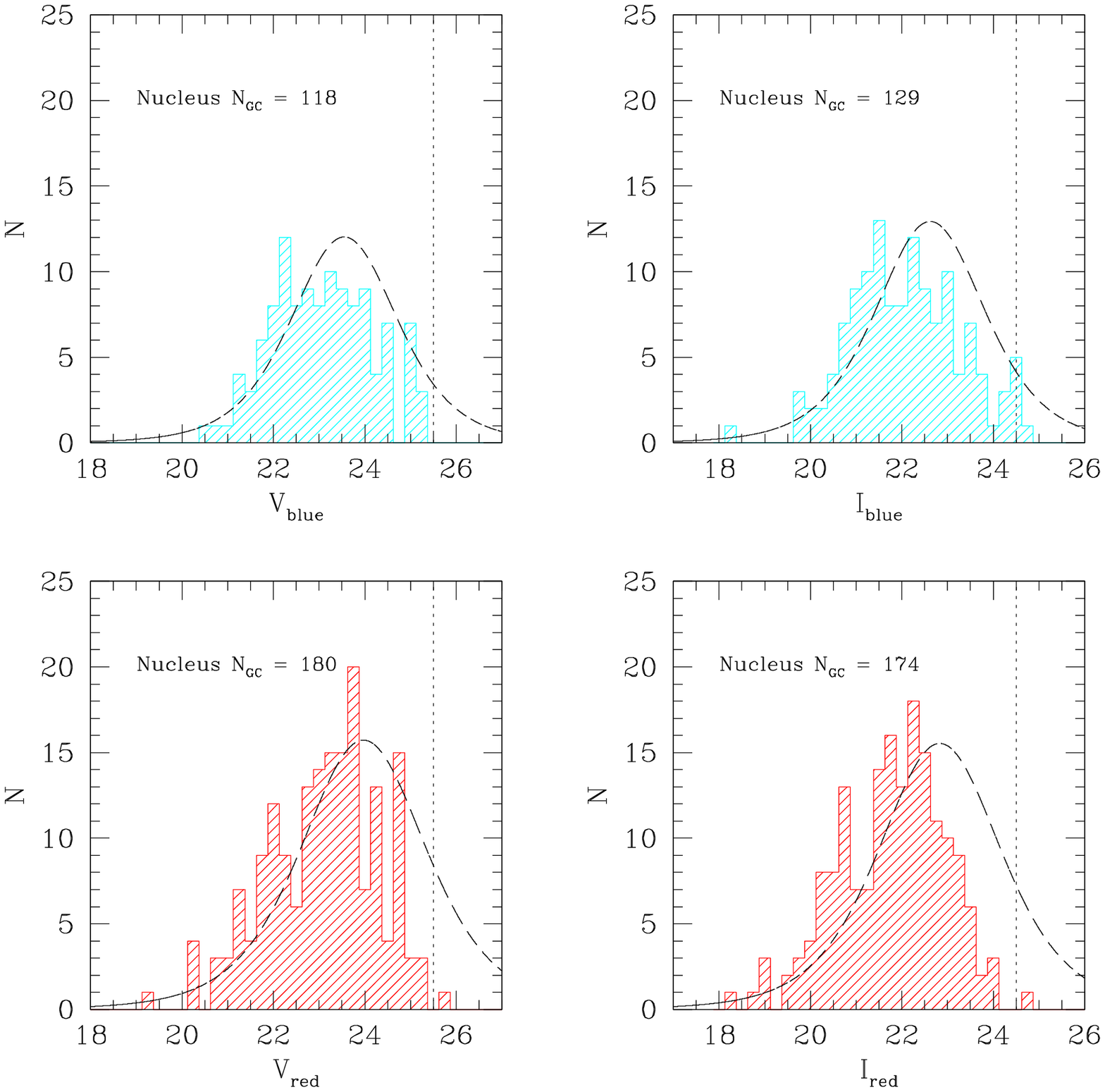,height=8cm,width=8cm
    ,bbllx=8mm,bblly=57mm,bburx=205mm,bbury=245mm}
\caption{Globular cluster luminosity functions of the halo (Brodie)
  pointings (upper 4 plots) and central (Westphal) pointing (lower 4
  plots). The number of clusters that went into the histograms up to
  the faintest completeness limit are indicated. The histograms are
  shown for the selected globular clusters, uncorrected for
  completeness and background variations (see text). The dashed curves
  are the best fitting $t_5$ functions as derived from our
  maximum--likelihood analysis, taking into account completeness and
  background corrections. The $t_5$ functions were arbitrarily
  normalized. The vertical dotted lines mark our 50\% completeness
  limit at the lowest background value.}
\label{ps:gclf}
\end{figure}

The Maximum--Likelihood routine of Drenkhahn (1999) was used to
determine the TO magnitudes. Our completeness functions fit down to
the 50\% completeness levels ($V_{\rm faintest}\simeq 25.5$, $I_{\rm
  faintest}\simeq 24.5$) were used as the input. The data were
convolved with the errors as a function of magnitude (see
Eq.~\ref{eq:Phi}). The input globular cluster data were selected as
described in Sect.~\ref{selection}. From KMM mixture modeling we
determined that $(V-I)=1.1$ is the best value for separating blue and
red clusters. For the analysis we split the full dataset into the
central (Westphal) pointing and the combination of the halo (Brodie)
pointings, since these two datasets had slightly different exposure
times, i.e.~slightly different limiting magnitudes/completenesses. The
TO magnitudes for the individual pointings are given in Table 3,
together with an average for the whole sample. The globular cluster
luminosity functions are shown in Fig.~\ref{ps:gclf}. The values in
Table 3 still need to be corrected for Galactic extinction
($A_V=0.069$ and $A_I=0.040$) and the additional aperture correction
(see Sect.~\ref{extended}), however, these corrections do not affect
the {\it relative} difference between the TO magnitudes.

We varied the input completeness and noise limits to the Maximum
Likelihood code, the resulting TO magnitudes all lie within $<0.15$
mag.

As a consistency check, we attempted to reproduce the mean
(uncorrected) colors of the sub--populations ($V-I$=0.99$\pm0.01$ and
1.24$\pm0.01$) by computing the differences between the $V$ and $I$ TO
magnitudes for a given population. We obtained $(V-I)_{\rm
  TO}=1.14\pm0.11$ for the blue population and $(V-I)_{\rm
  TO}=1.23\pm0.16$ for the red population, where the errors are the
errors of the $V$ and $I$ TOs taken in quadrature. The color for the
blue population derived in this way is somewhat redder than that
derived from the color distribution but the values agree to within the
errors for both populations.

\subsubsection{Monte--Carlo simulations}

The error in the TO magnitude directly affects the error in the
estimation of the age difference between the two sub--populations (see
Sect.~\ref{models}). We carried out Monte Carlo simulations to check
whether our relatively modest sample sizes (which were as low as 118
to 232 objects when divided into inner and outer samples) affect the
results significantly and to establish the associated error.

The blue and red samples were simulated in the $V$ filter using the
number of objects in each sample and assuming a Gaussian distribution
with the parameters derived above. These artificial samples were input
into the Maximum--Likelihood code. The 68\% confidence levels
($1\sigma$) on the TO magnitudes, from $10^4$ iterations, were 0.10
and 0.13 for the blue and red samples, respectively. These are in good
agreement with the $1\sigma$ errors returned by the
Maximum--Likelihood estimation.

\subsection{Comparison with previous observations}

The derived colors and TO magnitudes were compared to previous results
from other groups.

\subsubsection{The mean colors}

Geisler et al.~(1996) derived metallicity peaks for the blue and red
sub--populations from Washington photometry. They found peaks at
[Fe/H]$=-1.3\pm 0.38$ dex and $-0.1\pm 0.38$ dex for the blue and red
sub--samples, respectively. Our colors, derived in
Sect.~\ref{meancol}, can also be translated into metallicities.
However, large errors are introduced by the particular choice of a
conversion relation of the $(V-I)$ color into metallicity. Using the
relation given by Kissler-Patig et al.~(1998a) ([Fe/H]$=-(4.50\pm0.30)
+ (3.27\pm0.32)\cdot (V-I)$), which is tuned to reflect the non-linear
behavior at high metallicity, we obtain metallicities of
[Fe/H]$=-1.45\pm0.43$ dex and $-0.64\pm0.49$ dex. These values are
somewhat lower than, but within the errors of, the values derived by
Geisler et al.~(1996). Using the transformation of Kundu \& Whitmore
(1998), [Fe/H] $=-5.89+4.72\cdot (V-I)$, our data yield metallicities
of [Fe/H] $=-1.50\pm 0.05$ dex and [Fe/H] $=-0.32\pm 0.05$ dex. The
error corresponds to our photometric error in $(V-I)$ only. No
calibration error of Kundu \& Whitmore is considered. The metallicity
calibration of Couture et al.~(1990), [Fe/H] $=-6.096+5.05\cdot
(V-I)$, gives the somewhat larger values of [Fe/H] $=-1.39\pm 0.05$
dex and $-0.11\pm 0.05$ dex, similar to those derived by Geisler et
al.~(1996).

\subsubsection{The TO magnitudes}

TO magnitudes have been derived by several groups but only for the
whole sample (i.e.~an average of the red and blue populations) and in
different filters. Previous results should be compared to the average
of our blue and red sample, with the necessary correction for
reddening and aperture size: $V_0=23.76\pm0.20$ (including the error
on the TO magnitude and the error on the aperture correction).
Furthermore, all other groups assumed E$(B-V)=0.00$ towards NGC 4472
(using Burstein \& Heiles 1982/84) while we adopted the Schlegel et
al.~(1998) map from DIRBE/IRAS whose zero-point is offset by 0.020 in
E$(B-V)$. Our value would translate into $V_0=23.83\pm0.20$ if we
assumed E$(B-V)=0.00$.

Using the Washington--Johnson transformation of Geisler (1996),
$V\simeq T_1+0.5$, Lee et al.'s (1998) result translates into
$V_0=23.81\pm 0.07$, in excellent agreement with our result. We note
that Lee et al.~investigated their blue and red samples separately,
but did not find any significant difference in their TO magnitudes to
within $\simeq$~0.15 mag. However, their analysis was less rigorous in
this respect than ours, and their data suffered from more background
contamination, as is expected for ground--based observations.

Harris et al.~(1991) derived $B_{TO}=24.78\pm0.22$ with a 3--parameter
Gaussian fit to the GCLF. Using $B-V\simeq0.8$ for the total globular
clusters population, this translates into $V_{TO}=23.98\pm0.22$. Ajhar
et al.~(1994) derived $R_{TO}=23.3\pm0.2$ (read off their Fig.~14)
which, for $V-R\simeq 0.5$, translates into $V_{TO}=23.8\pm0.2$. These
results are in very good agreement with our determination.

\subsection{Radial Dependencies}
\label{radial}
\subsubsection{TO magnitudes as a function of radius}

\begin{figure}
  \psfig{figure=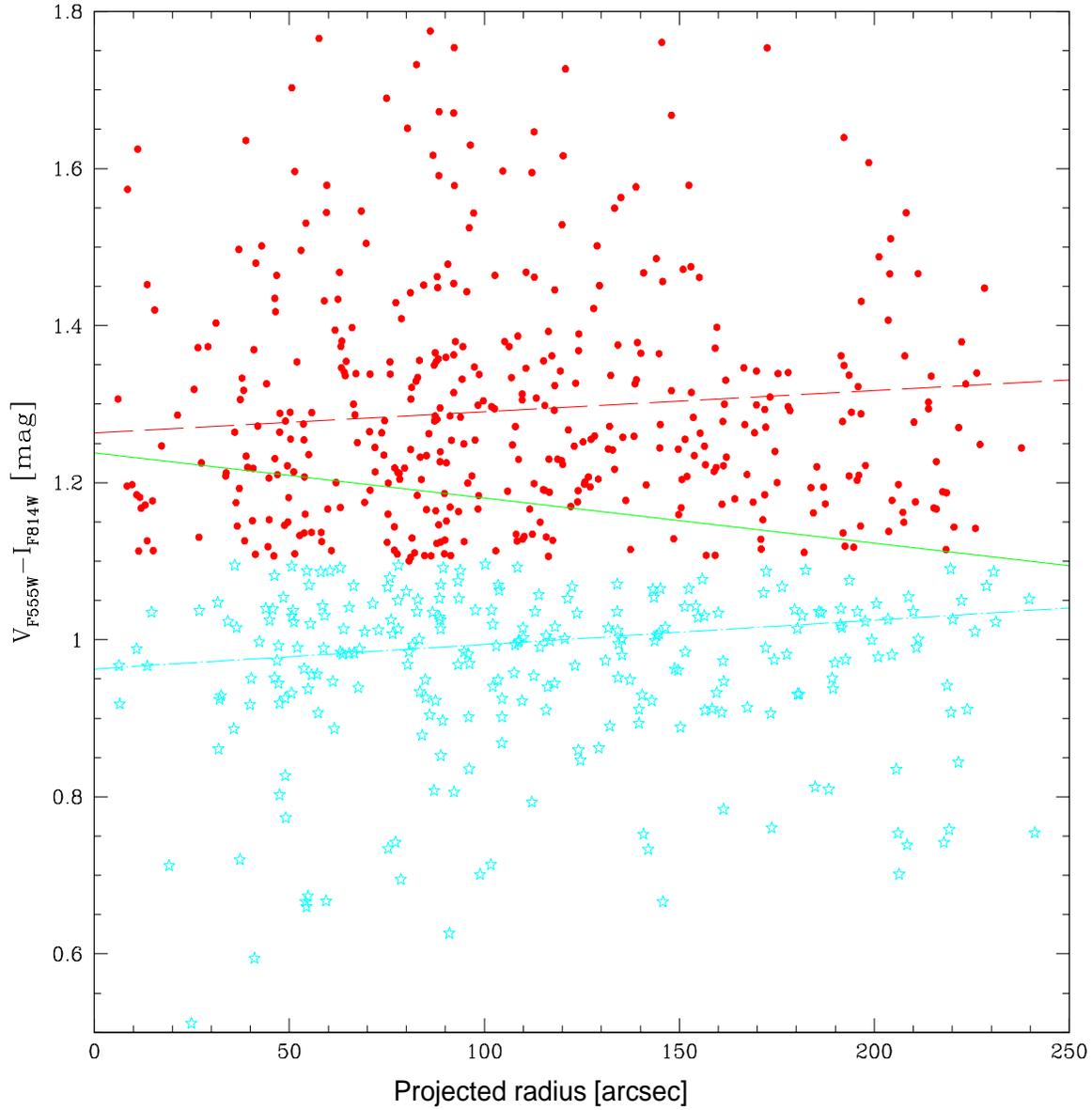,height=16cm,width=16cm
    ,bbllx=8mm,bblly=57mm,bburx=205mm,bbury=245mm}
\caption{Radial distributions of the red (open circles) and
  blue (stars) globular clusters. The dashed line represents the best
  weighted linear least--square fit to the red population.  The
  dot--dashed line gives the best fit to the blue population. The
  solid line shows the best fit for the entire globular cluster
  population. The gradient is twice as large as for each
  sub--population.}
\label{ps:colorrad}
\end{figure}

Variation in the TO magnitude with radius within a population can
occur if the mean age, metallicity or mass changes with radius. Large
age and metallicity gradients are ruled out by the absence of any
significant color gradient in the sub--populations (see
Fig.~\ref{ps:colorrad}). A systematic change in the mean mass with
radius could be caused by destruction processes shaping the globular
cluster mass function. These are more important towards the center of
the galaxy (e.g.~recent work by Vesperini 1997 and Gnedin \& Ostriker
1997), and significant changes are only expected within $\simeq 5$ kpc
of the center ($\simeq 65$\arcsec\ at 16 Mpc). There the peak of the
GCLF is expected to be brighter by up to $\simeq 0.3$ magnitudes, and
the peak should be sharper as a consequence of the preferential
destruction of low--mass clusters. Our inner sample (Westphal data)
includes clusters at projected radii between 2\arcsec\ and 127\arcsec\ 
with a mean of 65\arcsec\, and our outer sample (Brodie data) includes
clusters at projected radii between 76\arcsec\ and 252\arcsec\ with a
mean of 160\arcsec. Differences are therefore expected between these
two samples. For the blue population, the TO magnitude varies from the
inner to the outer regions by $0.12\pm0.20$ and $-0.29\pm0.25$ in $V$
and $I$ respectively. The weighted mean is $-0.04\pm0.16$, i.e.~we
observed no brightening of the TO towards the center, although the
uncertainty is on the order of the expected variation. For the red
population the TO magnitude varies from the inner to the outer region
by $0.33\pm0.17$ and $0.14\pm0.25$ in $V$ and $I$ respectively.
Formally, the TO appears brighter by $0.27\pm0.14$ (weighted mean) for
the inner red population. We do not see any changes in the dispersion
of the GCLF to within the errors.

This result is probably not yet secure enough to be worth commenting
on in depth. However, a difference between the red and blue population
could occur, for example, if the blue and red clusters are on
different orbits. Sharples et al.~(1998) found tentative evidence for
the blue clusters in NGC~4472 having a higher velocity dispersion and
higher rotation than the red ones. Kissler-Patig \& Gebhardt (1998)
found a similar (but firmer) result for M~87, another giant elliptical
that hosts enough globular clusters for such an analysis. If indeed
the blue clusters are preferentially on tangential orbits, while the
red clusters are preferentially on radial orbits, destruction
processes would be more efficient for the red population.

Finally, we note that Harris et al.~(1998) and Kundu et al.~(1999)
also looked for this effect in M~87. Kundu at al.~even looked for
trends in the blue and red populations separately. Neither group found
any brighting of the GCLF towards the center of M~87, but note again
that the uncertainties in the measurements are still of the order of
the expected signal.

\subsubsection{Color as a function of radius}
\label{colorrad}
We looked for color gradients within the sub--populations using a
weighted least squares fitting routine. In addition to the color cuts
($0.5<V-I<1.1$ and $1.1<V-I<1.8$), the samples were cut at $V$ = 24.8
to provide at least 50\% completeness even in the central regions. The
results are plotted in Fig.~\ref{ps:colorrad}, together with a fit for
the total sample. The derived gradients are $\Delta (V-I)/\Delta \log
r = 0.015\pm0.004$ mag/log(arcsec) and $0.013\pm 0.007$
mag/log(arcsec) for the blue and red population respectively.
Different color and magnitude cuts scatter these results within 0.01
mag/log(arcsec). Any positive gradients in the sub--populations are
either very small or non--existent. For the whole sample, however, we
find a clear negative gradient of $\Delta (V-I)/\Delta \log r =
-0.029\pm0.07$ mag/log(arcsec), corresponding to a metallicity
gradient of $\Delta$[Fe/H]$/\Delta \log r \simeq -0.15\pm0.02$
dex/log(arcsec) (somewhat dependent on the color--metallicity relation
used).

We confirm Geisler et al.'s (1996) result, that this gradient is
caused by the changing ratio of red to blue clusters as a function of
radius, rather than an overall decrease in metallicity with radius.
Quantitatively, we find a somewhat smaller gradient than reported by
Geisler et al.~(1996), who found $\Delta$[Fe/H]$/\Delta \log r \simeq
-0.4$ dex/log(arcsec). This is probably due to our smaller radial
coverage and our use of a less metallicity--sensitive color. Indeed,
our sample extends to only 250\arcsec\ while Geisler et al.'s data
extend out to $\simeq$ 500\arcsec. The ratio of blue to red clusters
varies by only $\simeq$ 10\% from our inner to outer regions, as
derived from a comparison of their cumulative radial distributions.

A similar situation is seen in M~87, for which Harris et al.~(1998)
report no color gradient within 60\arcsec , and Kundu et al.~(1999)
report a weak gradient of $\Delta (V-I)/\Delta \log r =
-0.017\pm0.012$ mag/log(arcsec) within 100\arcsec . However, Lee \&
Geisler (1993) report a strong gradient ($\Delta$[Fe/H]$/\Delta \log r
= -0.65\pm0.17$ dex/log(arcsec)) over 500\arcsec\ radius.

\subsection{Globular Cluster Sizes}
\label{gcsizes}
\begin{figure}
  \psfig{figure=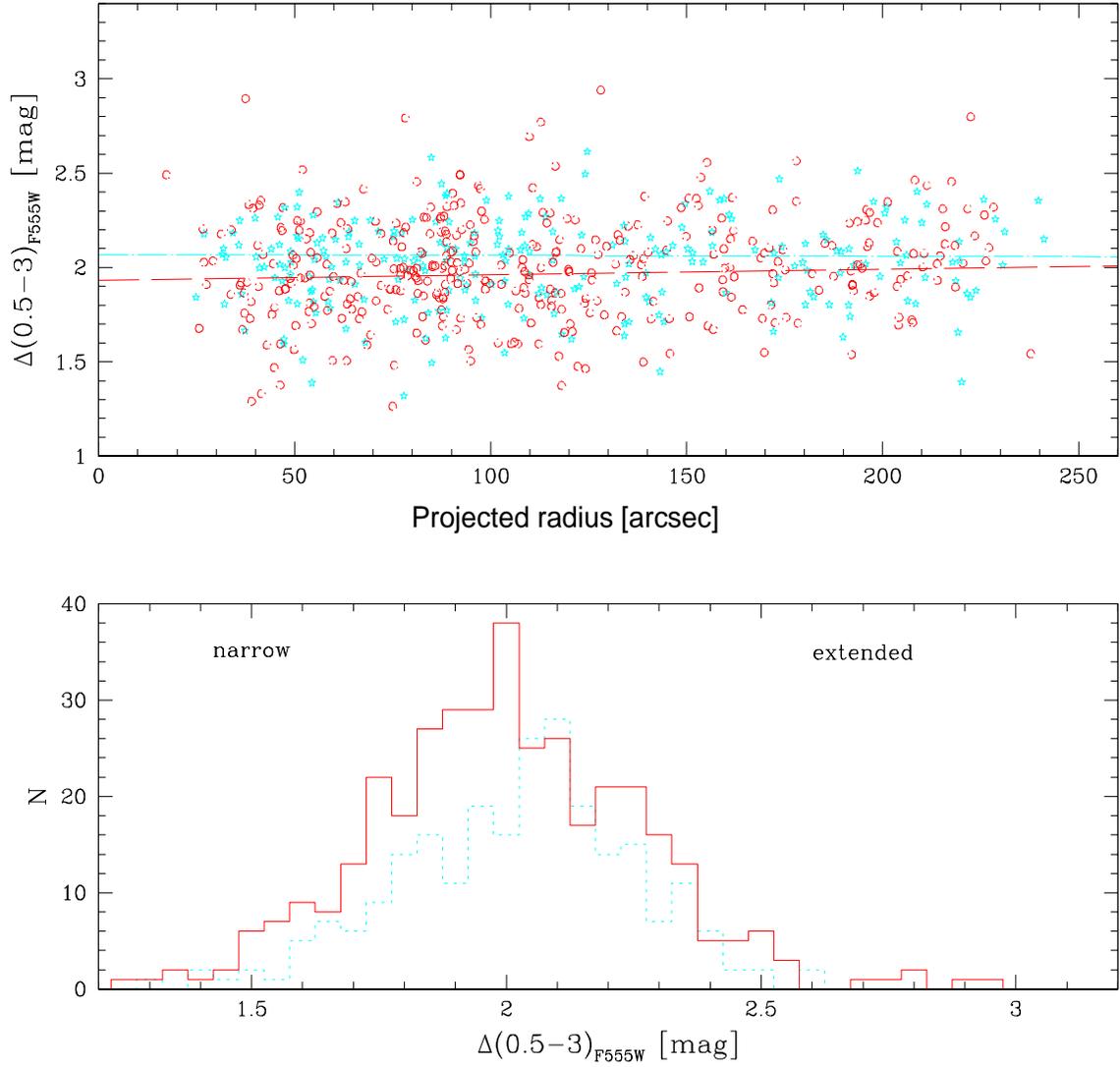,height=16cm,width=16cm
    ,bbllx=8mm,bblly=57mm,bburx=205mm,bbury=245mm}
\caption{Upper panel: Radial globular cluster size distribution 
  for blue (open stars) and red (open circles) globular clusters. The
  dot--dashed line is the best least--square fit to the blue
  sub--population while the dashed line gives the best fit to the red
  sub--population.
  Lower panel: Histogram of relative globular cluster sizes for the
  red population (solid histogram) and the blue population (dashed
  histogram). The sizes of the blue globular clusters are
  substantially larger than those of the red.}
\label{ps:sizerad}
\end{figure}

The high resolution of HST allows us to study the {\it relative} sizes
(see Sect.~\ref{extended}) of the globular cluster candidates using
the parameter $\Delta(0.5-3)$ (see Sect.~\ref{selection}). For
extended objects the magnitude difference in two different apertures
will appear large as the light profile contributes a non--negligible
amount of light to the outer aperture. The opposite is the case if the
light profile is narrow. For this analysis, we restricted ourselves to
clusters found on the Wide Field (WF) camera chips, to avoid the
problematic conversion between the WF and the Planetary Camera (PC).
Fig.~\ref{ps:sizerad} presents the histogram of globular cluster sizes
for both the blue and red sub--populations. The median values for
these two distributions point to a systematic size difference between
the blue and the red globular clusters. The median values for the blue
and red populations are $m_\Delta=2.06\pm 0.01$ mag and
$m_\Delta=2.02\pm 0.01$ mag, respectively, indicating that the blue
globular clusters are substantially larger than the red globular
clusters. A Maximum--Likelihood estimation of the mean and its error
yields the same results.

Interestingly, in both NGC~3115 (Kundu \& Whitmore 1998) and M~87
(Kundu et al.~1999), the blue clusters also appear significantly
larger than the red ones. The authors suggest this is due to different
radial distributions for the red and blue clusters, noting that
Galactic globular clusters with larger galactocentric distances have
larger half--light radii (van den Bergh 1994).

To study the radial dependence of globular cluster size, we plotted
the size parameter ($\Delta(0.5-3)$) versus radius and found no
gradients: $\Delta(0.5-3)=2.07(\pm0.03) - 0.5(\pm3.2)\cdot
10^{-4}\cdot r$ where $r$ is the radius in arcsec, and
$\Delta(0.5-3)=1.93(\pm0.03) - 2.9(\pm2.7)\cdot 10^{-4}\cdot r$, for
the blue and red clusters respectively (see also
Fig.~\ref{ps:sizerad}). The difference between blue and red clusters
is present at all radii. We therefore speculate that the different
sizes are a relic of the different formation processes, unless the red
and blue clusters are on significantly different orbits. For example,
the red clusters could be preferentially on radial orbits and be
systematically more affected than the blue clusters by the central
potential during their close passage near the galaxy center.

\section{The age difference between the main sub--populations}
\label{agediffs}
\subsection{Comparison with the models}
\label{models}
We compare our results (TO magnitude and mean color) to various
population synthesis models in order to derive the age {\it
  difference} between the blue and red globular cluster
sub--populations.

\begin{figure}
  \psfig{figure=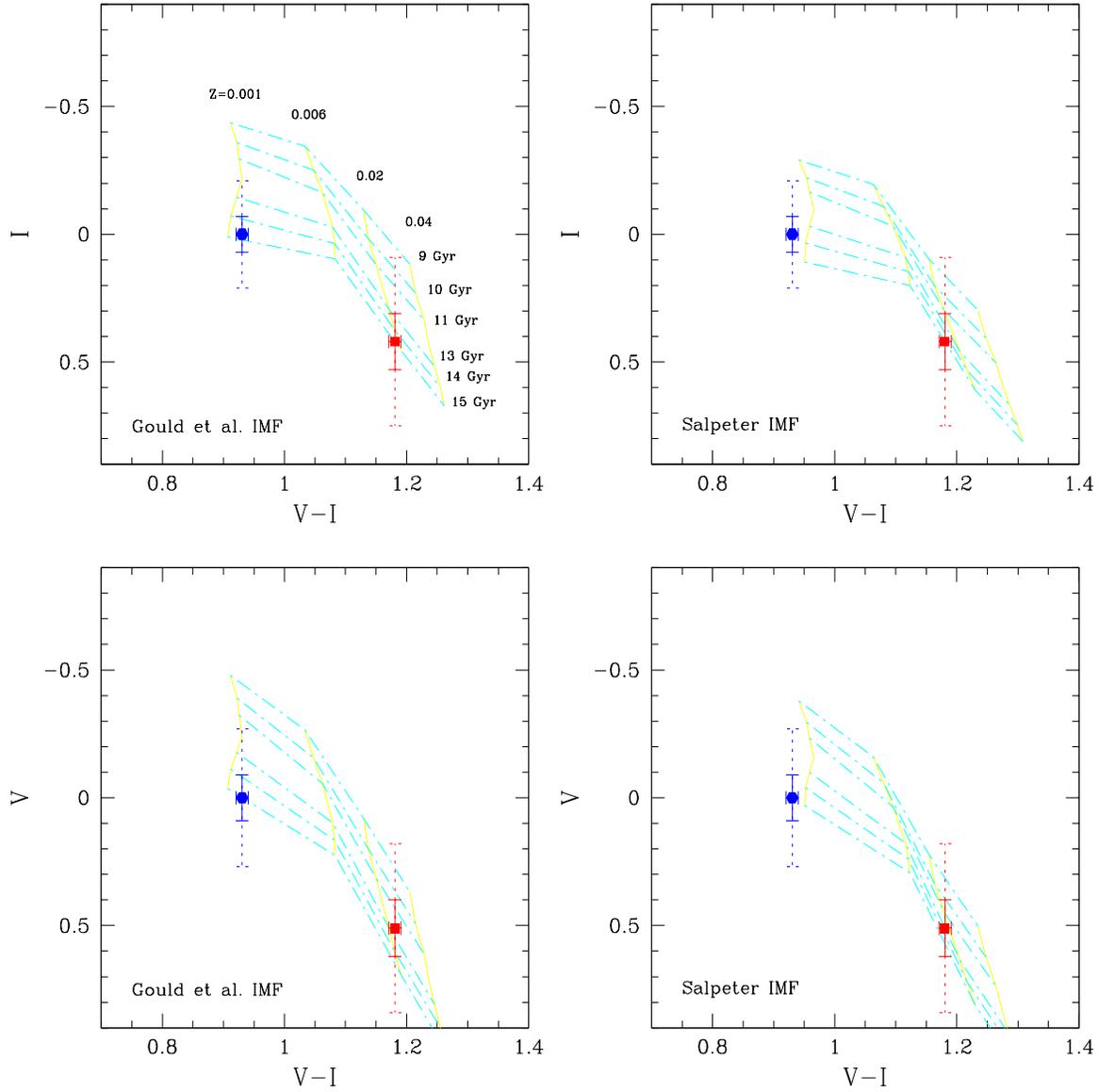,height=16cm,width=16cm
    ,bbllx=8mm,bblly=57mm,bburx=205mm,bbury=245mm}
\caption{Maraston's (1998) Simple Stellar Population models. 
  Upper panels: I--band data. Lower panels: V--band data. The solid
  grid lines represent iso--metallicity tracks of $Z$ = 0.001, 0.006,
  0.02, 0.04. Dash--dotted lines are isochrones of 9, 10, 11, 13, 14
  and 15 Gyr. Solid error bars represent 1$\sigma$ errors and dashed
  error bars refer to 3$\sigma$ errors. The averaged maximum age
  difference (age$_{\rm blue} -$ age$_{\rm red}$) is $\Delta T=0.7\pm
  1.8$ Gyr.}
\label{ps:mar}
\end{figure}

\begin{figure}
  \psfig{figure=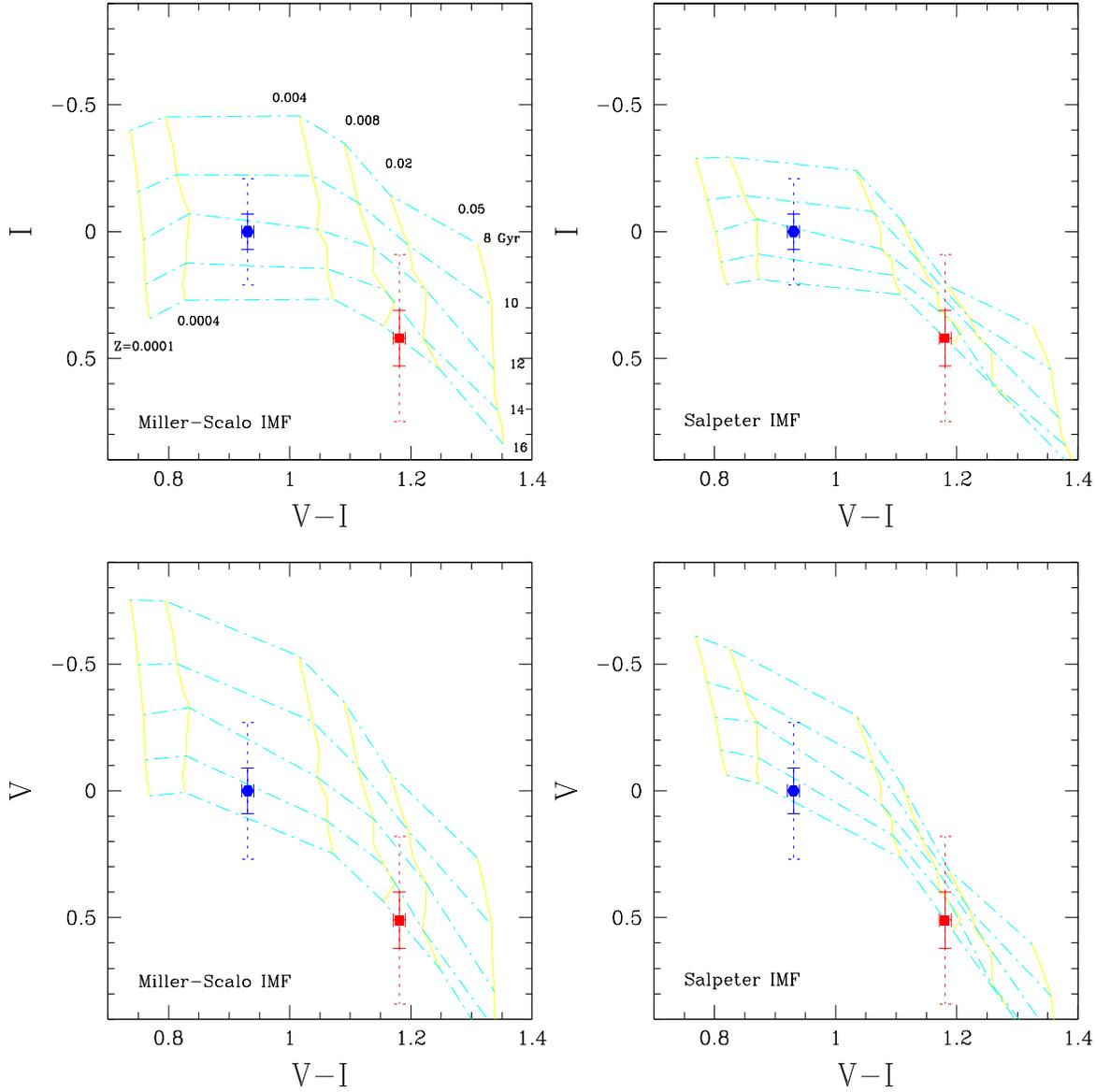,height=16cm,width=16cm
    ,bbllx=8mm,bblly=57mm,bburx=205mm,bbury=245mm}
\caption{Simple Stellar Population models by Kurth et al. (1999)
  calculated for a single stellar burst of $10^7$ Gyr duration. Upper
  panels: I--band TO magnitudes. Lower panels: V--band TO magnitudes.
  Within the grids the solid lines (vertical) represent
  iso--metallicity tracks of $Z$ = 0.0001, 0.0004, 0.004, 0.008, 0.02,
  0.05 (from left to right) which correspond to [Fe/H]$=-2.3, -1.7,
  -0.7, -0.4,$ 0.0 and 0.4. The dash--dotted (horizontal) lines are
  isochrones of 8, 10, 12, 14 and 16 Gyr. The solid error bars
  represent the 1$\sigma$ errors. The dotted error bars give the
  3$\sigma$ deviations. An averaged maximum age difference of $\Delta
  T=-2.3\pm 1.6$ Gyr is derived.}
\label{ps:kurth}
\end{figure}

\begin{figure}
  \psfig{figure=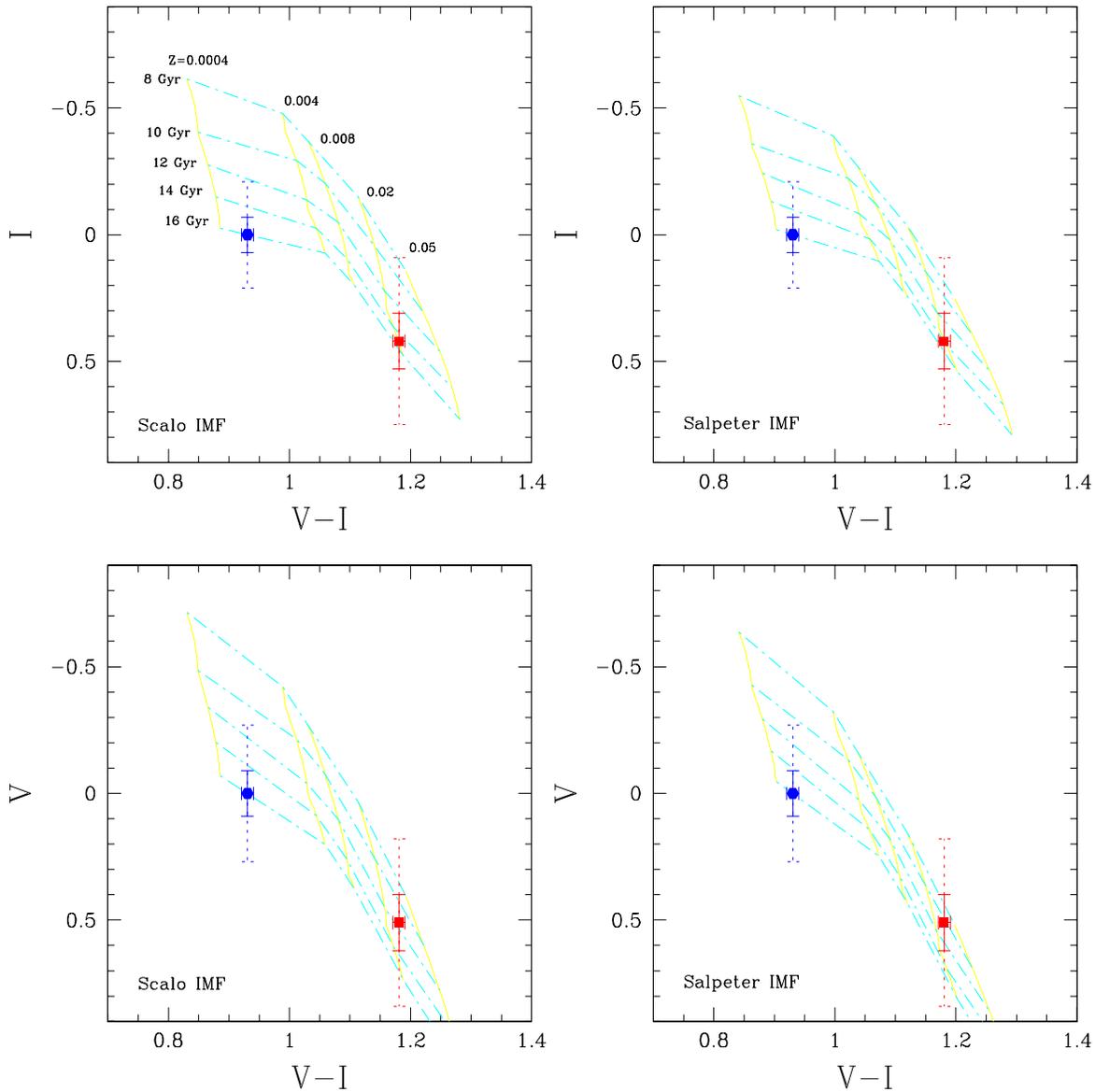,height=16cm,width=16cm
    ,bbllx=8mm,bblly=57mm,bburx=205mm,bbury=245mm}
\caption{Simple Stellar Population models by Bruzual \& Charlot
  (1996). Upper panels: I--band data. Lower panels: V--band data. The
  solid grid lines represent iso--metallicity tracks of $Z$ = 0.0004,
  0.004, 0.008, 0.02 and 0.05. Dash--dotted lines are isochrones of 8,
  10, 12, 14, 16 Gyr. Using these grids an averaged maximum--age
  difference of $\Delta T=3.3\pm 2.2$ Gyr was derived.}
\label{ps:bc}
\end{figure}

\begin{figure}
  \psfig{figure=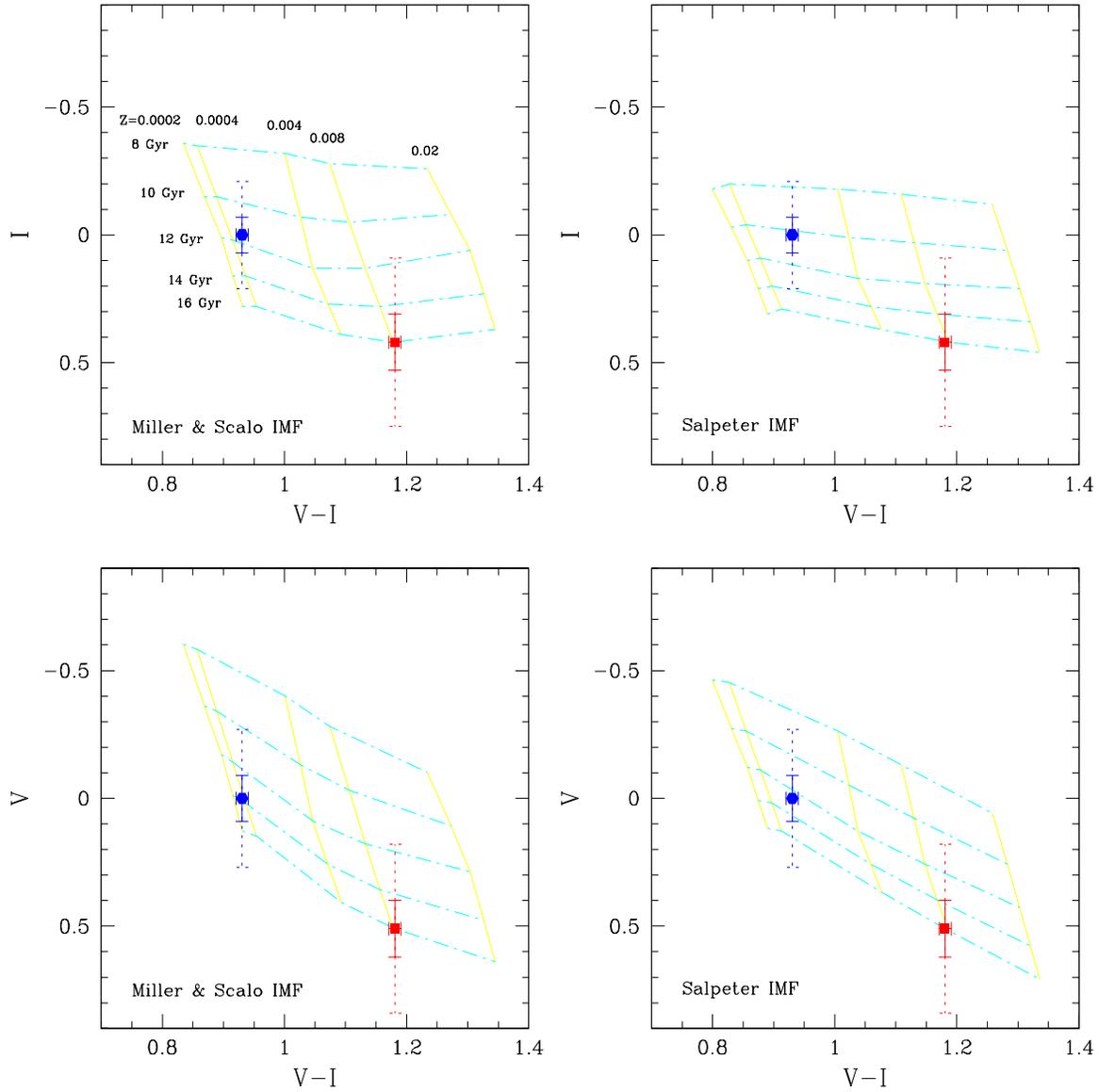,height=16cm,width=16cm
    ,bbllx=8mm,bblly=57mm,bburx=205mm,bbury=245mm}
\caption{Worthey's (1994) Simple Stellar Population models.
  Upper panels: I-band data. Lower panels: V-band data. The solid grid
  lines represent iso--metallicity tracks of $Z$ = 0.0002, 0.0004,
  0.004, 0.008, and 0.02. Dash--dotted lines are isochrones of 8, 10,
  12, 14, 16 Gyr. The averaged maximum age difference is $\Delta
  T=-4.0\pm 1.6$ Gyr.}
\label{ps:wo}
\end{figure}

We used the new models from Maraston (1998) as well as the most recent
models from the G\"ottingen Group (Kurth et al.~1999), and the models
from Bruzual \& Charlot (1996) and Worthey (1994). In each case we
compared our data to every available Initial Mass Function (IMF),
i.e.~usually the Salpeter IMF and a multi--slope IMF. The full range
of metallicities was used as well as the necessary age range. Since we
are only interested in age {\it differences} the exact normalization
of the absolute magnitude, which is dependent on mass and
mass--to--light (M/L) ratio for a given IMF, is not important. The
grids in Fig.~\ref{ps:mar} to \ref{ps:wo} can, therefore, be somewhat
arbitrarily shifted in the y--direction. In the following, we fixed
the grids in the y--direction such that the oldest sub--population
lies on the oldest computed isochrone (15 Gyr or 16 Gyr). Further, we
arbitrarily set the blue population to $V=0$ and $I=0$ in order to
compare the different models directly. There is no freedom in the
x--direction since the color is not mass--dependent, and the grid is
fixed in $V-I$ for a given stellar population.

For the mean colors of the blue and red sub--populations we used the
colors derived in Section \ref{meancol}. The TO values used were the
ones derived for the full sample (see Table \ref{tab:magcol}). We
stress again that additive corrections to both the blue and red TO
magnitudes (i.e.~reddening correction, potential systematic errors,
aperture corrections) only influence the difference as second order
effects, if at all, and are not a concern.

However, we make two important underlying assumptions. We assume that
the M/L ratios for blue and red clusters are only dependent on age and
metallicity (i.e.~the IMFs of the blue and red clusters are similar).
Second, we assume that blue and red cluster populations have the same
underlying mass distribution (by which we mean globular cluster
masses, not IMFs). Under these assumptions, TO magnitudes are
influenced only by age and metallicity. We briefly discuss the
validity of these assumptions and the errors they potentially
introduce.

\subsubsection{The M/L and mass distribution assumptions}
\label{ml}
Kissler-Patig et al.~(1998b) review the empirical reasons for
expecting the mass distributions and M/L ratios to be similar among
various sub--populations. Briefly, the M/L assumption is essentially
the assumption that both blue and red clusters have similar IMFs, at
least below 1 $M_\odot$ where it influences our result. Observations
so far for globular clusters in the Local Group seem to confirm this
picture (e.g.~Dubath \& Grillmair 1997). In the case of the mass
distribution, globular cluster formation seems to follow a universal
mechanism (e.g.~Elmegreen \& Efremov 1997). The constancy of the TO
magnitude from galaxy to galaxy (e.g.~the review by Whitmore 1997) and
the observed mass distributions of young clusters (e.g.~brief review
by Schweizer 1997) support this picture.

The possibility that size differences might indicate different
formation processes for blue and red clusters was discussed in Section
\ref{gcsizes}. It is not clear, however, whether it is only the mean
sizes or also the mean masses that are affected. Moreover, our
assumption would fail if, despite similar mass distributions at
formation, blue and red clusters are affected differently by
destruction processes. There is a hint (see Section \ref{radial}) that
this might be the case in the inner region, where the difference
between the blue and the red TO magnitudes is smaller by $0.32\pm0.24$
(combining the $V$ and $I$ information) than in the outer region. (We
note for completeness that an age gradient in the population would
have the same observable effect as a mass gradient). If this effect is
real, the TO difference of the total sample, as used, could be
affected by a 0.1 to 0.2 mag systematic error. We estimate that this
empirical determination gives about the right order of magnitude for
the potential systematic error in our method and, in what follows, we
will assume a potential systematic error of $\simeq 0.2$ mag, which
translates into roughly 4 Gyr.

Note that for similar M/L values of the blue and red clusters, such a
difference of 0.2 mag would translate into only a 20\% difference in
the mass characterizing the TO. Irrespective of the accuracy needed
for our study, blue and red clusters {\it have} roughly similar mass
distributions.

\subsubsection{Avoiding assumptions about the masses}

For completeness, we mention a method for checking the validity of
these assumptions and a method for avoiding them.

For the chosen filters, age and characteristic mass influence the TO
far more than metallicity, which is fairly well determined by the
color. Therefore, one could determine the mean mass difference (and
check the above assumption) by obtaining an independent estimate of
the mean age difference. An independent age estimate could be obtained
with spectroscopy of a large number of globular clusters and the
measurement of an age--sensitive absorption line index. Several tens
of H$\beta$ measurements in each sub--population would allow the mean
H$\beta$ value to be determined to within $<0.1$ \AA\ which would
translate into an accuracy in the mean age difference of $\simeq 2$
Gyr. This in turn would allow us to verify that the characteristic TO
masses of the red and blue populations lie within 10\% to 20\% of each
other.

An alternative method for determining the mean age difference without
making any assumptions about the mass distributions would be to use
mass--indepen\-dent quantities that depend only on age and
metallicity. Broad--band colors are such quantities. A plot similar to
Figures \ref{ps:mar} through \ref{ps:wo}, but in a color--color plane,
would remove any uncertainties due to different mass distributions.
However, it is impossible to find two colors in the optical and NIR
spectrum that disentangle age and metallicity as efficiently as the
color/TO magnitude combination.

\subsubsection{Comparison with Maraston (1998)}

In Figure \ref{ps:mar}, we compare our results to the models of
Maraston (1998). She computed single burst population models with a
Salpeter (1955) and multi-slope IMF (Gould et al.~1998) for
metallicities Z$=0.001, 0.006, 0.02$ and $0.04$, and ages up to 15
Gyr. We refer the reader to the original paper for further details of
the modeling.

The derived age difference, which is quoted in the following as
(age$_{\rm blue}$)$-$(age$_{\rm red}$), varies between $\sim -0.3$ Gyr
(Gould et al.~IMF, $I$ band) and $\sim +2.4$ Gyr (Gould et al.~IMF,
$V$ band), with errors of about 2.7 Gyr (1$\sigma$). Within the
errors, the blue and red population appear coeval, with a formal
difference of $+0.7\pm1.8$ Gyr (straight mean of the four computed
differences with their dispersion shown as the error). A 3$\sigma$
limit on the age difference would be around 6 Gyr.

\subsubsection{Comparison with Kurth et al.~(1999)}
\label{kurth}
In Figure \ref{ps:kurth}, we compare our results to the recent models
of Kurth et al.~(1999). They computed single burst models with a
Salpeter (1955) and a Miller--Scalo (1979) IMF for 6 metallicities
between Z$=0.0001$ and $0.05$, and ages up to 16 Gyr.

The age difference varies between $-3.8$ Gyr (Sal\-pe\-ter IMF, $I$
band) and $-0.4$ Gyr (Salpeter IMF, $V$ band) with a typical error of
2.2 Gyr. The mean difference is $-2.3\pm1.6$ Gyr, i.e.~the red
population appears somewhat older than the blue one. The 3$\sigma$
differences are around $-7$ and $+2.5$ Gyr.

\subsubsection{Comparison with Bruzual \& Charlot (1996)}

In Figure \ref{ps:bc}, we compare our results to the models of Bruzual
\& Charlot (1996). We used the single burst populations computed with
a Salpeter (1955) IMF and a Scalo (1986) IMF, with metallicities
between Z$=0.0004$ and $0.05$. Their models with a Miller--Scalo
(1979) IMF were not used. The tracks for their highest metallicity
(Z$=0.1$) are not shown. The oldest modeled populations is 16 Gyr.

The age differences vary between $+1$ Gyr (Scalo IMF, $I$ band) and
$+6$ Gyr (Salpeter IMF, $V$ band), with an typical error of 2.8 Gyr.
The mean difference is $+3.3\pm2.2$ Gyr. The 3$\sigma$ limits lie
around $-3$ and $+10$ Gyr.

\subsubsection{Comparison with Worthey (1994)}

In Figure \ref{ps:wo}, we compare our results to Worthey's (1994)
single burst models. He computed models with a Salpeter (1955) IMF and
a Miller--Scalo (1985) IMF for metallicities between Z$=0.0002$ and
0.05, up to an age of 16 Gyr.

The age difference varies between $-5.8$ Gyr (Salpeter IMF, $I$ band)
and $-2.1$ Gyr (Miller--Scalo IMF, $V$ band), with a typical error of
2.5 Gyr. The mean difference is $-4.0\pm1.6$ Gyr, i.e.~the red
populations appears older than the blue one. The 3$\sigma$ limits are
around $-9$ and $+1$ Gyr.

\subsubsection{The maximum age difference between the globular
  cluster populations}

We note that the $I$ band results give a systematically smaller
difference (in the sense we defined it: age$_{\rm blue}-$age$_{\rm
  red}$) than the $V$ band results. Except for the Bruzual \& Charlot
(1996) models, the results from the $I$ band suggest that the red
population is marginally older than the blue one.

The different models predict mean differences anywhere between $-4.0$
Gyr and $+3.3$ Gyr, with a mean of $-0.6$ Gyr and a dispersion of
$3.2$ Gyr. It would be somewhat arbitrary to assign more weight to one
model than to another, so that we are currently limited in the age
determination by the model uncertainties. Nevertheless, it seems that
both populations are coeval within the errors, and we can exclude at
the 99\% confidence level one population being half as old as the
other.

If our results are affected by a systematic error, e.g.~if the smaller
sizes of the red globular clusters do indeed translate into a lower
``mean'' mass (such that the red TO magnitude is $\simeq 0.2$ lower --
see Section \ref{ml}), the age difference would be affected by $\simeq
+4$ Gyr.


\subsection{Implications for the star formation episodes of NGC 4472}
\label{discussion}
The above results suggest that the vast majority of globular clusters
and stars formed at high redshift in this cluster galaxy. As already
mentioned in Section \ref{intro}, several other lines of evidence lead
to similar results for cluster galaxies. Interestingly, the globular
cluster results imply that two major star--formation events took place
at high redshift. Those two events appear to have happened moderately
closely in time but from differently enriched gas. The blue clusters
appear to have formed from gas with less than $1/20$th solar
metallicity while the red clusters formed out of gas with close to
solar metallicity. Furthermore, the mean sizes of the blue and red
clusters appear to be significantly different. This suggests different
conditions at the formation epoch, unless they are directly related to
the metallicity and associated cooling processes in the molecular
clouds in which the clusters originated.

It is unclear whether the blue or the red clusters are the older ones.
At face value it seems odd that the more metal--rich clusters would be
the older ones. However, in scenarios where the blue clusters formed
in halo clouds (e.g.~Kissler-Patig et al.~1998b) or were accreted with
dwarf galaxies (e.g.~C\^ot\'e et al.~1998, Hilker et al.~1999a,b),
this could be understood in terms of the large structures
(e.g.~bulges) collapsing first, followed shortly thereafter by the
smaller structures (clouds/dwarf galaxies). This, however, would force
a revision of the merger scenario which clearly predicts that the red
clusters will be younger than the blue ones.

Alternatively, if two sub--populations of clusters are coeval or if
the blue clusters are the older ones, age differences of a few Gyr
between the different sub--populations could be explained in all
globular cluster/galaxy formation scenarios, i.e.~the above--mentioned
scenarios, merging (Ashman \& Zepf 1992), or in--situ formation of
distinct populations (e.g.~Forbes et al.~1997, Harris et al.~1998,
Harris et al.~1999). We note only that, in the merger scenario,
merging would have to have taken place at early times ($z>1$ at the
3$\sigma$ level) which is somewhat at odds with the recent theoretical
predictions from hierarchical clustering models (e.g.~Kauffmann 1996,
Baugh et al.~1998). These tend to predict that the main star formation
in cluster environments occurred at redshifts $z<1$.


\section{A distance estimation for NGC 4472}
\label{distance}

The TO magnitude ($V_{TO}$) of the GCLF appears to be essentially
universal and, therefore, can be used as a distance indicator
(e.g.~the review by Whitmore, 1997). $V_{TO}$ depends on the age and
metallicity of the globular clusters (Ashman, Conti \& Zepf 1995, see
also Sect.~\ref{models}), and weakly on galaxy type (Harris 1997)
which may be a consequence of the age/metallicity dependence. The
derived TO magnitude for the globular clusters in NGC 4472 can be used
to obtain a distance to the galaxy.

\subsection{The absolute turn--over magnitude}

The method is best calibrated using the TO magnitude of the Milky Way
GCLF, since it involves the minimum number of steps on the distance
ladder.  The M31 GCLF serves as a strong check on this calibration
(cf.~Ferrarese et al.~2000; Kavelaars et al.~1999).

The Milky Way system is dominated by old globular clusters of mean
metallicity [Fe/H]$\simeq-1.4$. It is, therefore, well--suited to
fitting the TO magnitude of the blue population in NGC~4472 (see
Sect.~\ref{models}) and no correction for age--metallicity dependence
need be applied. However, we have to assume that age--metallicity
dominates any dependence on galaxy type, and that globular cluster
mass functions are intrinsically similar, which seems justified to
first order (e.g.~Kissler-Patig et al.~1998b).

Della Valle et al.~(1998) recently re--derived $V_{TO}$ for the Milky
Way system. To derive individual distances to all Galactic globular
clusters, they adopted a horizontal branch (HB) magnitude --
metallicity relation derived from globular cluster main sequence
fitting using distance measurements to sub--dwarfs from {\tt
  HIPPARCOS} (in this case by Gratton et al.~1997). From the resulting
GCLF they obtained $V_{TO}=-7.62\pm0.06$ noting that, depending on the
adopted HB magnitude--metallicity relation, this result could suffer
from a 0 to $-0.2$ mag systematic error.

Drenkhahn \& Richtler (1999) used a different approach. They assumed
the LMC distance to be the fundamental zero--point (here
$(m-M)=18.46\pm0.06$), and derived an RR Lyrae magnitude--metallicity
relation. The remaining steps were similar to those of Della Valle et
al.. Drenkhahn \& Richtler found $V_{TO}=-7.61 \pm 0.08$ and
$I_{TO}=-8.48 \pm 0.10$ for the Milky Way GCLF.

\subsection{The distance to NGC 4472}

The apparent TO magnitude derived for the blue population is
$V_{TO}=23.62\pm0.09$ and $I_{TO}=22.48\pm0.07$ (see
Tab.~\ref{tab:magcol}). These values need to be corrected for the
photometric term discussed in Sect.~\ref{extended} ($\Delta
C_V=0.050\pm0.015$, $\Delta C_I=0.080\pm0.010$), as well as for
Galactic extinction ($A_V=0.069$, $A_I=0.040$). Using the absolute TO
magnitudes discussed above, we derive distance moduli of
$(m-M)_V=31.11\pm0.19$ and $(m-M)_I=30.84\pm0.14$ (adding all errors
in quadrature), i.e.~a mean $(m-M)=30.99\pm0.11$ (where the error does
not include any possible systematic error in the calibration). The
corresponding distance is $15.8\pm0.8$ Mpc.

A quick comparison with distances derived for NGC~4472 by other
methods shows very good agreement (despite completely different
calibrations). Jacoby et al.~(1990) derived $(m-M)=30.84\pm0.11$ from
the planetary nebulae luminosity function. Tonry et al.~(1990) and
Neilsen (1999) derived $(m-M)=30.78\pm0.07$ and $(m-M)=30.93\pm0.08$,
respectively, from surface brightness fluctuations.

Cepheid distances now exist for 7 spiral galaxies in the Virgo Cluster
(cf.~Pierce et al.~1994 and Ferrarese et al.~2000). With the exception
of NGC~4639 (Saha et al.~1996), which appears to be background to the
main body of spirals, the mean distance to the other 6 galaxies is
$(m-M)=31.01\pm0.07$ or 16.0$\pm$0.6 Mpc. Our GCLF distance to
NGC~4472 is in excellent agreement with this distance, which supports
the view that the Virgo cluster spirals and ellipticals are at
essentially the same distance.

\section{Summary}
\label{summary}
The age difference between the two major globular cluster
sub--populations in an early--type galaxy has been determined. We have
developed an age differentiating method which uses the precise TO
magnitudes and colors of the sub--populations. We have applied this
method to NGC~4472, matching our observables to several SSP synthesis
models.

An extended--object analysis showed that at the distance of Virgo
(i.e.~$\sim$16 Mpc) HST is capable of resolving single globular
clusters. Thus it is necessary to make an additive correction (of $C_V
= 0.050\pm0.015$ mag and $C_I = 0.080\pm0.010$ mag at the distance of
Virgo) to the mean correction of $C=0.1$ mag (for point sources) when
extrapolating HST photometry obtained in 0.5\arcsec\ radius apertures
to total luminosities. This corresponds to a color correction of
$C_{(V-I)}=0.03\pm0.018$.

We found the globular cluster color distribution in NGC~4472 to be
clearly bimodal (at the 99.99\% confidence level) with two peaks
located at $V-I=0.99\pm0.01$ and $1.24\pm0.01$ corresponding to
$(V-I)_0=0.93\pm0.02$ and $(V-I)_0=1.18\pm0.02$ when corrected for
reddening and finite aperture. We determined that the red globular
clusters have a color distribution almost twice as broad as the blue
ones.

We divided the whole globular cluster population into an inner and an
outer sample, as defined by the HST different pointings. We used an
improved Maximum--Likelihood estimator, which accounts for varying
background values, to determine the TO magnitudes for each
sub--population in each radial sample (see Tab.~\ref{tab:magcol}).
The TO of the blue sub--population is the same for the inner and outer
sample. By contrast, the red sub--population appears to be marginally
brighter by $0.27\pm0.14$ mag in the inner region than in the outer
region. This difference is of the expected order of magnitude if
destruction processes act on the red sample within 5 kpc.

The radial color distributions of each sub--population show almost no
gradients. However, we detected an overall gradient of $\Delta
(V-I)/\Delta\log r = -0.029\pm0.07$ mag/log(arcsec) in the system,
which is mostly an effect of the changing ratio of red to blue
globular clusters with radius (cf.~Geisler et al.~1996).

Taking advantage of HST's high spatial resolution, we estimated the
mean sizes of the globular clusters using the $\Delta(0.5-3)$
parameter. The blue globular clusters were found to be larger than the
red ones, as was already observed in M~87 (Kundu et al.~1998) and
NGC~3115 (Kundu \& Whitmore 1997). The most likely explanations are
that the size difference reflects different formation processes, or
that red clusters are on preferentially radial orbits, while blue
clusters are on preferentially tangential orbits.

We used 4 different SSP models to derive the mean age difference
between the red and blue sub--population. Our underlying assumption
was that both blue and red populations have similar mass distributions
which appears to be valid at least at the 10\% to 20\% level (which,
nevertheless, corresponds to a potential systematic error of $\sim 4$
Gyr). The mean age difference between the blue (metal--poor) and the
red (metal--rich) sub--populations is age$_{blue} -$ age$_{red} =
-0.6\pm3.2$ Gyr, where the error is the dispersion introduced by the
different models (for a given model the error on our method is $\sim$2
Gyr).

We conclude that the globular clusters (and by association the stars)
in NGC~4472 formed in two major episodes/mechanisms that were coeval
within a few Gyr.


The images as well as electronic lists of the photometry are available
from the authors. Also available is the maximum--likelihood code as an
{\tt IRAF} task used to derive the globular cluster luminosity
functions taking into account the background variations over the
field.


\subsubsection*{Acknowledgments}

We are very thankful to Claudia Maraston for customizing part of her
models for our purposes and providing electronic files, as well as to
Georg Drenkhahn for his help in implementing his Maximum Likelihood
code. We also thank Sandy Faber, Duncan Forbes, Karl Gebhardt, and
Klaas de Boer for interesting discussions.

THP gratefully acknowledges support from the UCO/LICK Observatory.
MKP was partly supported by the Alexander von Humboldt Foundation.
This research was funded by HST grants GO.05920.01-94A and
GO.06554.01-95A and faculty research funds from the University of
California at Santa Cruz. This research was partly supported by NATO
Collaborative Research grant CRG 971552.


\clearpage

\onecolumn

\clearpage 

\begin{deluxetable}{l ll ll}
\tablenum{1}
\tablecaption{General properties of NGC 4472}
\tablehead{
\colhead{NGC 4472} & \colhead{} & \colhead{} 
}
\startdata
& RA(2000)$^a$         & $ $12h 29m 46.79s $\pm1.25$s \\
& DEC(2000)$^a$        & $+$08$^o$ 00\arcmin\ 01.50\arcsec\ $\pm1.25$\arcsec\ \\
& $m_V$ $^b$           & $8.41\pm0.06$ mag \\
& $V-I$ $^c$           & $1.24\pm0.01$ mag \\
& $B-V$ $^b$           & $0.96\pm 0.01$ mag \\
& MType $^b$           & E2/S0(2) \\
& $V_{\rm helio}$ $^b$ & $868\pm8$ km s$^{-1}$\\
& Distance $^d$        & $15.8\pm 0.8$ Mpc \\
\enddata
\label{tab:ngc4472}
\tablenotetext{}{a: Laurent-Muehleisen S.A., et al.~1997}
\tablenotetext{}{b: RC3v9 (de Vaucouleurs, et al.~1991)}
\tablenotetext{}{c: Poulain P., 1988}
\tablenotetext{}{d: this paper}
\end{deluxetable}

\begin{deluxetable}{l ll ll}
\tablenum{2}
\tablecaption{Summary of the observations}
\tablehead{
\colhead{Prg. ID + PI} & \colhead{RA(2000)} & \colhead{DEC(2000)} & 
\colhead{F555W exp. time$^a$} & \colhead{F814W exp. time$^a$} 
}
\startdata
GO.5236 Westphal       & 12 29 48.73 & $+$07 59 55.85 & 1800 sec & 1800 sec \\
GO.5920 Brodie (north) & 12 29 45.22 & $+$08 02 34.38 & 2200 sec & 2300 sec \\
GO.5920 Brodie (south) & 12 29 45.20 & $+$07 57 30.11 & 2200 sec & 2300 sec \\
\enddata
\label{tab:observation}
\tablenotetext{a}{Effective exposure time, after images were stacked.}
\end{deluxetable}

\begin{deluxetable}{ll ll ll l}
\tablenum{3}
\tablecaption{Turn--over magnitudes and dispersions for each
  sub--population luminosity function}
\tablehead{
\colhead{Filter} & \colhead{Population} & \colhead{$M_{TO}$} & 
\colhead{$\Delta M_{TO}$} & \colhead{$\sigma_{t_5}$} & 
\colhead{$\Delta\sigma_{t_5}$} & \colhead{$N_{\rm GC}$}
}
\startdata
Total sample & & & & & & \\
& & & & & & \\
$V$   & blue & 23.62 & 0.09  & & \\
      & red  & 24.13 & 0.11  & & \\
$I$   & blue & 22.48 & 0.07  & & \\
      & red  & 22.90 & 0.11  & & \\
& & & & & & \\
Westphal (inner) sample & & & & & & \\
& & & & & & \\
$V$   & blue & 23.55 & 0.15 & 1.21 & 0.10 & 118 \\
      & red  & 23.97 & 0.19 & 1.42 & 0.10 & 180 \\
$I$   & blue & 22.62 & 0.21 & 1.24 & 0.14 & 129 \\
      & red  & 22.83 & 0.19 & 1.39 & 0.10 & 174 \\
& & & & & & \\
Brodie (outer) sample & & & & & & \\
& & & & & & \\
$V$   & blue & 23.68 & 0.11 & 1.36 & 0.09 & 175 \\
      & red  & 24.30 & 0.13 & 1.39 & 0.09 & 232 \\
$I$   & blue & 22.33 & 0.08 & 1.14 & 0.07 & 178 \\
      & red  & 22.97 & 0.13 & 1.39 & 0.08 & 213 \\
& & & & & & \\
\enddata
\label{tab:magcol}
\tablenotetext{}{The TO magnitudes are neither corrected for
  reddening, nor for the aperture correction discussed in
  Sect.~\ref{extended}. The corrections for galactic reddening are
  $A_V=0.069$ and $A_I=0.040$. The corrections for aperture, 
    in addition to the correction of Holtzman $C=0.1$ (which is already
    included) are $C_V=0.050\pm0.015$ and $C_I=0.080\pm0.010$ (which are
    not included here).}
\end{deluxetable}

\clearpage


\clearpage

\end{document}